\documentclass{aa} 
\usepackage{graphicx}
\usepackage[dvipsnames]{xcolor}
\usepackage{tabularx}
\usepackage{booktabs}
\usepackage{txfonts}
\usepackage{natbib}
\bibpunct{(}{)}{;}{a}{}{,}  %to follow the A&A style

\mathchardef\period=\mathcode`.
\DeclareMathSymbol{.}{\mathord}{letters}{"3B}

\def\kms{{\rm km\, s^{-1}}}
\def\aap{{\em A}\&{\em A}}

\def\apjl{{\em ApJ}}
\def\apjs{{\em ApJS}}

\begin{document}

   \titlerunning{Revealing ringed galaxies in group environments}
   \authorrunning{Fernandez et al. 2023}
   \title{Revealing ringed galaxies in group environments}

%   \subtitle{}

   \author{Julia Fernandez\inst{1},
          Sol Alonso\inst{1},
                  Valeria Mesa\inst{2,3,4},
        Fernanda Duplancic\inst{1}
          }

   \institute{Departamento de Geof\'{i}sica y Astronom\'{i}a, CONICET, Facultad de Ciencias Exactas, F\'{i}sicas y Naturales, Universidad Nacional de San Juan, Av. Ignacio de la Roza 590 (O), J5402DCS, Rivadavia, San Juan, Argentina\\
              \email{fernandezmjulia@unsj-cuim.edu.ar }
         \and  Instituto de Investigación Multidisciplinar en Ciencia y Tecnología, Universidad de La Serena, Raúl Bitrán 1305, La Serena, Chile.
         \and Association of Universities for Research in Astronomy (AURA)
\and 
Grupo de Astrofísica Extragaláctica-IANIGLA, CONICET, Universidad Nacional de Cuyo (UNCuyo), Gobierno de Mendoza
 }
             
   \date{Received xxx; accepted xxx}

   \abstract
   {} 
   {We explore galaxies with ringed structures inhabiting poor and rich groups with the aim of assessing the effects of local density  environments on ringed galaxy properties.
   }
{We identified galaxies with inner, outer, nuclear, inner+outer (inner and outer rings combined), and partial rings that reside in groups by cross-correlating a sample of ringed galaxies with a group catalog obtained from Sloan Digital Sky Survey (SDSS).
The resulting sample was divided based on group richness, with groups having $3 \leq  N_{rich} \leq  10$ members classified as poor, while groups having $11 \leq  N_{rich} \leq  50$ were classified as rich.
To quantify the effects of rings and the role of local density environment on galaxy properties, we constructed a suitable control sample for each catalog of ringed galaxies in poor and rich groups, consisting of non-ringed galaxies with similar values for the redshift, magnitude, morphology, group masses, and environmental density distributions as those of ringed ones.
We explored the occurrence of ringed galaxies in poor and rich groups and analyzed several galaxy properties, such as star formation activity, stellar populations, and colors, with respect to the corresponding comparison samples.}
{We obtained a sample of 637 ringed galaxies residing in groups. We found that about 76\% of these galaxies inhabit poor groups, whereas only about 24\% are present in rich groups. Inner rings are prevalent in both rich and poor groups, while nuclear rings are the least common in both groups. Regarding the control sample, about 81\% galaxies are found in poor groups and about 19\% in rich ones. We find that the percentages of ringed galaxies with bar structures are similar, regardless of whether the group is  rich or poor.
In addition, ringed galaxies inhabiting groups display a reduction in their star formation activity and aged stellar populations, compared to non-ringed ones in the corresponding control samples. However, the star formation rate is higher for nuclear rings in poor groups than for other types. This disparity may stem from the environmental influence on the internal processes of galaxies, either enhancing or diminishing star formation.
Ringed galaxies also show an excess of red colors and tend to populate the green valley and the red sequence of color-magnitude and color-color diagrams, with a surplus of galaxies in the red sequence, while non-ringed galaxies are found in the green valley and the blue region. These trends are more significant in galaxies with ringed structures residing in rich groups. %Our findings provide valuable insights into the relationship between ringed structures and their surrounding environments, paving the way for further explorations in this area of study.
}
   {}


\maketitle 
%
%________________________________________________________________

\section{Introduction}

Galactic rings are elliptical or circular structures formed by gaseous and stellar components that are believed to be directly related to the dynamics of their host spiral galaxies (\citealt{Kormedy1979}; \citealt{Diaz2019}). About one-fifth of all disk galaxies present a ring pattern and, additionally, one-third have partial rings \citep[also known as pseudo-rings;][]{Buta1996}.

Ringed galaxies have been classified by several authors and two major classes are generally recognized. Gas accumulation at certain resonances due to the action of gravity torques by galactic bars produces normal resonant rings, which make up the majority of the observable rings in spiral galaxies. The nuclear, inner, and outer rings are associated with the main Lindblad resonances \citep[]{Buta2017}. Based on visual analysis, resonance ring galaxies have been classified as R(S) by \cite{deVaucouleurs1959} and as O-type by \cite{Few1986}. On the other hand, a clear relationship exists between the formation rate of rings and galaxy minor interactions (\citealt{Donghia2008}; \citealt{Tous2023}). Hence, minor mergers produce catastrophic rings and can result in three different types of rings (accretion rings, polar rings, and collisional rings),  depending on the mass, gas content of the satellite galaxy, and its orbit around the more massive spiral galaxy \citep{Schweizer1987,Lynds1976, Donghia2008,smir22}. Furthermore, \cite{Helmi2003} found that the disruption of satellite galaxies in orbits that lie almost in the same plane as the disk leads to the formation of ring-shaped stellar structures. However, numerical simulations have shown that the probability of ring formation during close passages of galaxies is small \citep{Tutukov2016}. This second type was classified as RING by \cite{deVaucouleurs1976} and as a P-type by \cite{Few1986} (see \citealt{Elagali2018}; \citealt{Lamb1993}).

Different observational analysis of the properties of particular ringed galaxies have been made. \cite{Gusev2003} found a  typical scenario of star formation activity and colors in the ringed barred galaxy NGC 2336, taking its morphological type into account. Similarly, \cite{Grouchy2010} found that barred and non-barred ringed galaxies exhibit similar star formation rates (SFR) and show no dependence on the presence of a ring structure in the disk.
\cite{Silchenko2018} studied the star formation in two S0 galaxies with outer rings, NGC 6534 and MCG 11-22-015, finding that the outer gas sources feeding star formation may be a consequence of tidal harassment of the neighbor by NGC 6534 and a minor merger for MCG 11-22-015, whilst the star formation histories in the rings are different.
More recently, \cite{Fernandez2021} (hereafter F21) presented a statistical analysis of various properties of a sample of face-on ringed galaxies, comparing them to a suitable control sample of non-ringed galaxies. According to their findings, ringed galaxies consistently exhibit a lower efficiency in star formation activity, producing new stars at a slower rate compared to non-ringed ones. Additionally, galaxies with ringed structures show an excess of redder colors and higher metallicity values compared to  galaxies without rings.

It is widely known that galaxies in dense environments, such as groups and clusters, exhibit different properties compared to their isolated counterparts (e.g.,  \citealt{Dressler1980},  \citealt{Balogh_2004},  \citealt{Baldry2006}, \citealt{Skibba2009}). There is clear evidence to suggest that the properties of galaxies, such as luminosity, color, and stellar mass, vary substantially with the environment in which they reside (\citealt{Lietzen2012}, \citealt{Allington1993}) whereas other properties, such as structure, are only indirectly correlated with the environment (e.g., \citealt{Kauffmann2004}, \citealt{Blanton2005}). For example, one of the most fundamental correlation between the morphological types of galaxies and the environment in the local Universe is called the morphology-density relation (\citealt{Oemler1974}, \citealt{Dressler1980}). This relationship shows that star-forming galaxies dominated  by disks reside in regions of the Universe with lower density than quiescent elliptical galaxies. {On larger scales, which represent the environment of the supercluster-void network, the morphology-density relation was first discovered by \cite{Einasto1987}. Similarly, in \cite{Balogh_2004} analyzed the connection between color, luminosity and environment. They proposed that the rate at which a star-forming galaxy evolves is primarily determined by its intrinsic properties. They also suggested that any environmentally induced transformation from a blue to red color must occur rapidly or at high redshifts.

Regarding the dependence between ringed galaxies and density environment, different studies show contradictory results. In their study of the local morphology-environment relation $(z\sim0)$ for bright galaxies, \cite{Wilman2012} found no clear dependence between the frequency of inner and outer rings and the local environment. Additionally, \cite{Buta2019} examined the optical morphologies of isolated galaxies in the  Analysis of the Interstellar Medium of Isolated Galaxies (AMIGA) sample, finding no correlation between the detection of inner and outer rings or of galactic bars and the parameters associated with their isolation. This finding suggests two possible scenarios. Firstly, it implies that the impact of the environment on the formation of these features in galaxies is minimal. Alternatively, it raises the possibility that the measurement of the isolation parameters exhibits a significant level of variation, making it difficult to differentiate among subtly different environments that may or may not trigger the formation of bars and rings in isolated galaxies. The researchers also considered another explanation for the presence of bars and rings in isolated galaxies. They proposed that these features might have originated from past interactions, which would indicate that these characteristics have been sustained for several Gyr. However, other authors such as \cite{Madore1980}, who analyzed a sample of nearby spirals with companion galaxies, found that rings are more common in systems with fewer companions than average. Furthermore, \cite{Elmegreen1992} studied the influence of environment on galaxies with outer and pseudo-rings finding that the fraction of SB0 and SB0/a galaxies with outer rings is less frequent in dense environments, while pseudo-rings increase in abundance with increasing environmental density. The fraction of late-type non-barred galaxies with outer pseudo-rings also decreases with density.

Motivated by these findings and with the purpose of contributing to the knowledge on this topic, our study aims to delve into the impact of dense environments on the characteristics exhibited by ringed galaxies.
To this end, we analyze spiral galaxies with and without rings inhabiting poor and rich groups to assess the influence of the environment in altering ringed galaxy properties, such as star formation rate, age of the stellar population, and colors. We use homogeneous and statistically complete catalogs; specifically, we use our sample of ringed galaxies (see F21) and the catalog of groups constructed by \cite{Tempel2017} from the Sloan Digital Sky Survey (SDSS).

This paper is structured as follows. Section 2 describes the database, the process of selecting ringed galaxies within groups, and the occurrence of different types of rings in galaxies within groups. Section 3 presents the criteria utilized for constructing reliable control samples. In Sections 4 and 5, we analyze the frequency of bars in ringed galaxies and their distribution within groups. Section 6 is focused on examining the characteristics of ringed galaxies in poor and rich groups, with an emphasis on their star formation activity, stellar populations, and color indexes in comparison to non-ringed galaxies. We also investigate the correlation of these traits with group-centric distance and local environmental density.
Finally, Section 7 provides a summary of the main conclusions drawn from our analysis.
The adopted cosmology throughout this paper is $\Omega_m=0.3$, $\Omega_{\Lambda}=0.7$, and $H_0=70 \kms \rm Mpc ^{-1}$.

%__________________________________________________________________

\section{Ringed galaxies in groups}

\begin{figure*}
\centering
\includegraphics[width=0.95\textwidth]{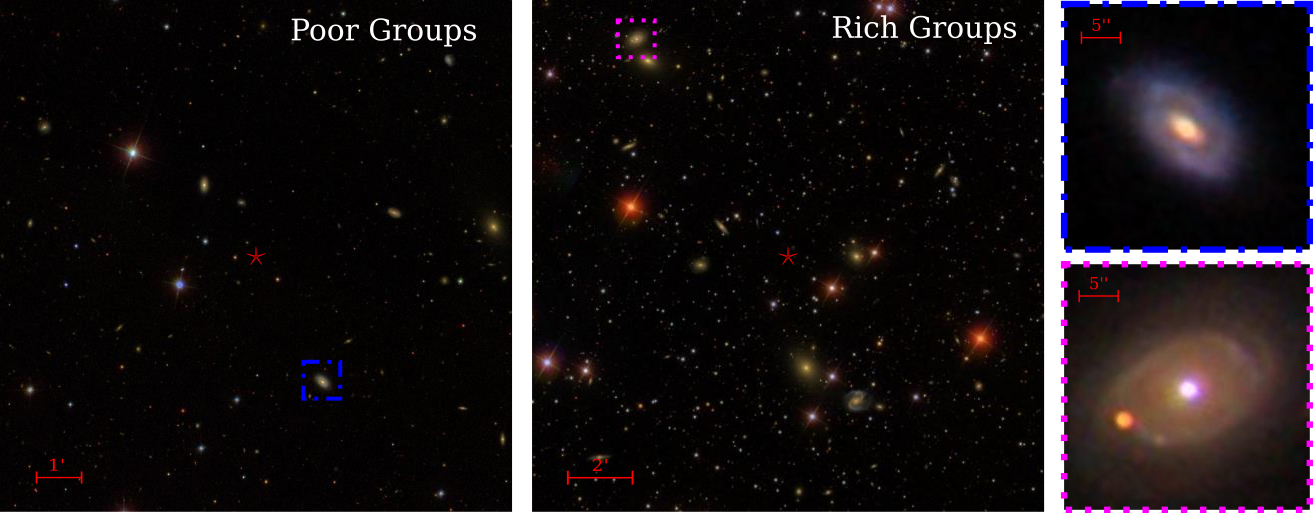} 
\caption{Typical examples of ringed galaxies in poor (left) and rich (right) groups are shown. In the right area of this figure, we present the zoomed images of outer ring galaxies in these groups. The red star represents the position of the group center. In all images, north is up and east is left.}
\label{eg}
\end{figure*}

This work is based on spectroscopic and photometric data from the Sloan Digital Sky Survey Data Release 14 (SDSS-DR14; \citealt{Abolfathi2018}).
DR14 is the second data release of the fourth phase of the SDSS \citep[SDSS-IV, 2014-2020;][]{sdssiv}, which includes all previous versions. It contains images, optical spectra, infrared spectra, integral field unit spectra, and catalog data, such as parameters measured from images and spectra.
For our study, we calculated k-corrections band-shifted to $z=0 \period1$ using the software \texttt{k-correct\_v4.2} by \cite{Blanton2007}. Furthermore, k-corrected absolute magnitudes were calculated from Petrosian apparent magnitudes converted to the AB system. We make use of the u, g and r-bands in the \textit{ugriz} system.

In our previous work (F21), we built a catalog of galaxies with ringed structures derived from the SDSS-DR14. This catalog will serve as one of the foundations for our present study.
By a detailed visual inspection of SDSS images, we classified 8529 face-on spiral galaxies (selecting objects with axial ratio $ b/a> 0 \period 5$, inclination $i<60^{\circ} $ and concentration index value,  C\footnote{$C=r90/r50$ is the ratio of Petrosian 90\%- 50\% r-band light radii} %\citep{Abraham1994}, 
$ <2 \period 8 $),  brighter than $g =$ 16.0 mag and in the redshift range 0.01 $<z<$ 0.1. Based on the different types of ringed structures, we obtained a catalog of 1868 ringed galaxies, representing a fraction of 22\% with respect to the full sample of spiral galaxies. 
The remaining 6661 galaxies were non-ringed.

For this study, we also used the catalog of groups of galaxies from the SDSS, as described by \cite{Tempel2017}.
This catalog is based on data from the SDSS DR12 (\citealt{Eisenstein2011}, \citealt{Alam2015}), where groups and clusters of galaxies were identified using a modified friends-of-friends (FoF) algorithm. The FoF method searches for galaxy pairs that are closer to one another than a given cut-off separation (or neighborhood radius). This algorithm was designed to find number overdensities in spectroscopic galaxy surveys and has also been modified to look for structures in simulated galaxy data sets (for further explanation of the method, see, e.g., \citealt{Huchra1982}, \citealt{Botzler2004}, \citealt{Tago2008}). Moreover, the FoF group membership is refined by multimodality analysis to find subgroups, where the virial radius of the group and escape velocity are used to expose unbound galaxies \citep{Tempel2016a,Tempel2017}. 
This catalog comprises 88662 groups with at least two members, of which 6873 systems have at least six members and 498 group mergers with up to six groups. 
In addition, this is a homogeneous catalog of groups where the primary parameters are independent of distance. As a result, projection effects are a more benign problem in the analysis. To explore the properties of ringed galaxies in group environments in comparison with galaxies without rings, we identified ringed galaxies that reside in rich and poor groups by cross-correlating the total ringed and non-ringed samples with galaxies in the group catalog described previously. 

Several authors \citep{Smith2022,Lacerna2022,Gozaliasl2018,vanderBurg2017} agree that for a system to fall into the group category, its halo masses must oscillate in a range of 13 $< Log (M_{200}/M_{sun}) < $ 14.5.
\cite{Li2019} used the relationship between the richness ($N_{rich}$) and virial mass ($M_{200}$) as a reference for the division of the sample of the poor and rich groups, finding that galaxy systems with halo masses between 13 $<Log (M_{200}/M_{sun})<$ 14.5 are found to have a richness ranging from 3 to 50.
Following the method of \cite{Li2019} to empirically define  galaxy systems with $3 \leq N_{rich} \leq 10$ as poor groups and those with $11 \leq N_{rich} \leq 50$ as rich groups, we divided the sample of ringed galaxies inhabiting  poor and rich groups into two subsamples according to their richness in the same way.

As a result of the previously mentioned process, we obtained a total sample of 637 ringed galaxies in groups. This value represents a fraction of 34.1\% with respect to the full sample of 1868 ringed galaxies. 
In addition, from the total sample of 637 ringed galaxies in high-density environments, $\approx$ 76\%  are located in poor groups, while only $\approx$ 24\% of these galaxies reside in rich groups, as can be observed in Table \ref{tab:Tabar}.

\begin{table} 
\center
\caption{Numbers and percentages (and their standard errors) of ringed and control galaxies in poor ($3 \leq N_{rich} \leq 10$) and rich groups ($11 \leq N_{rich} \leq 50$).}
\begin{tabular}{|c c c | }
\hline
Ringed Galaxies & Number & Percentage  \\
\hline
\textsc{Poor Groups} & 487 & 76.45 $\pm$ 0.87\% \\
\textsc{Rich Groups} &  150 & 23.55 $\pm$ 0.49\%  \\
\hline
\hline
Control Sample & Number & Percentage  \\
\hline
\textsc{Poor Groups} & 514 & 80.69 $\pm$ 0.89\% \\
\textsc{Rich Groups} &  123 & 19.31 $\pm$ 0.44\%  \\
\hline
\end{tabular}
{\small}
\label{tab:Tabar} 
\end{table}

\begin{table*} 
\center
\caption{Numbers and percentages (and their standard errors) of galaxies with different types of  rings in poor ($3 \leq N_{rich} \leq 10$), rich  ($11 \leq N_{rich} \leq 50$), and all groups.}
\begin{tabular}{|c|c c | c c|c c | }
\hline
 &  \multicolumn{2}{|c|}{Poor Groups} &   \multicolumn{2}{|c|}{Rich Groups} & \multicolumn{2}{|c|}{All Groups}   \\
\hline
Ring Type & Number & Percentage & Number & Percentage &  Number & Percentage \\
\hline
\hline
\textsc{inner ring} & 225 & 46.20 $\pm$ 0.68\% & 74 & 49.33 $\pm$ 0.70\% & 299 & 46.94 $\pm$ 0.68\% \\
\textsc{outer ring} &  45 & 9.24 $\pm$ 0.30\% & 25 & 16.67 $\pm$ 0.41\% & 70 & 10.99 $\pm$ 0.33\% \\
\textsc{i+o rings} & 86 & 17.66 $\pm$ 0.42\% & 22 & 14.67 $\pm$ 0.38\% & 108 & 16.95 $\pm$ 0.41\%  \\
\textsc{nuclear ring} &  34 & 6.98 $\pm$ 0.26\% & 8 & 5.33 $\pm$ 0.23\% & 42 & 6.59 $\pm$ 0.26\% \\
\textsc{partial ring} & 97 & 19.92 $\pm$ 0.45\% & 21 & 14.00 $\pm$ 0.37\% & 118 & 18.52 $\pm$ 0.43\%  \\
\hline
TOTAL  & 487 & 100\% & 150 & 100\% & 637 & 100\% \\ 
\hline
\end{tabular}
{\small}
\label{tab:Table2}
\end{table*}

Furthermore, we conducted an analysis of different types of ringed structures, including inner, outer, and inner+outer (cases in which the inner and outer rings coexist within the same galaxy are treated as a special class and do not contribute to other categories of individual rings, such as inner rings or outer rings), as well as nuclear and partial rings, to investigate their frequency in spiral galaxies as a function of local environments, taking into account poor and rich groups. Figure~\ref{eg} shows typical examples of outer ring galaxies in poor and rich groups selected from our sample.

It is necessary to emphasize that one of the features where resolution is important is in the detection of nuclear rings. In our study, the majority of nuclear rings are located within $\approx$ 2.2 kpc in radius. If any of the nuclear rings exceeded this limit, we followed an approach similar to that of \cite{Comeron2010}, which considered nuclear rings with radii above 2.0 kpc only if they were located within a bar or in unbarred galaxies that had inner, outer, and nuclear rings. In this manner, we aimed to mitigate any potential bias that these specific rings might introduce into the findings of our research.

In Table \ref{tab:Table2}, we display numbers and percentages of galaxies with different ring types found in poor and rich groups, revealing that $\approx$ 47\% of the ringed galaxies in group environments present inner rings, making them the most common type of ring in both rich and poor groups. 
In contrast, nuclear rings are the least frequent in both group samples. However, both types of rings do not show significant differences depending on the environment within the errors considered. Additionally, we can clearly observe that partial rings are more frequent in poor groups than in rich ones. The result that the occurrence of partial rings in ringed galaxies decreases as with increasing environmental density suggests that these features are not a result of tidal interactions with neighboring galaxies, but instead may be originated from internal processes within the galaxy and are later affected by the presence of tidal forces \citep{Buta1996}.

On the other hand, the fraction of galaxies with outer rings increases with group richness ($>$3$\sigma$ 
following Gaussian error propagation), contrasting the prevailing expectation that galaxies in poor groups are more likely to facilitate the maintenance of outer rings compared to galaxies in rich groups. Some studies in the literature have suggested that any significant tidal interaction with another galaxy could distort or destroy the outer ring \citep{Elmegreen1992}.

\begin{table*}[h!]
\centering
\caption{Number and percentages of galaxies exhibiting different ring types with barred and unbarred structures (regarding the total number of ringed galaxies within poor, rich, and all groups, respectively). Standard errors are also included.} 
\resizebox{\textwidth}{!}{
\begin{tabular}{|l|l|l|l|l|l|l|l|l|}
\hline
&\multicolumn{2}{l|}{Poor groups}&\multicolumn{2}{l|}{Rich groups}&\multicolumn{2}{l|}{All groups}\\
\hline
\hline
 Ring Type&Barred&Unbarred&Barred&Unbarred&Barred&Unbarred\\
\hline
\hline
\textsc{inner ring} &178 (36.55$\pm$ 0.60\%)&47 (9.65$\pm$ 0.31\%)&59 (39.33$\pm$ 0.63\%)&15 (10.00$\pm$ 0.32\%)&237 (37.21$\pm$ 0.61\%)&62 (9.73$\pm$ 0.32\%)\\ \hline
\textsc{outer ring} &5 (1.03$\pm$ 0.10\%)&40 (8.21$\pm$ 0.29\%)&3 (1.26$\pm$ 0.14\%)&22 (14.67$\pm$ 0.38\%)&8 (1.26$\pm$ 0.11\%)&62 (9.73$\pm$ 0.32\%)\\ \hline
\textsc{i+o rings}
&63 (12.94$\pm$ 0.36\%)&23 (4.72$\pm$ 0.22\%)&17 (11.33$\pm$ 0.34\%)&5 (3.33$\pm$ 0.18\%)&80 (12.56$\pm$ 0.35\%)&28 (4.40$\pm$ 0.21\%)\\ \hline
\textsc{nuclear ring} &1 (0.21$\pm$ 0.05\%)&33 (6.78$\pm$ 0.26\%)&0 (0\%)&8 (5.33$\pm$ 0.23\%)&1 (0.16$\pm$ 0.04\%)&41 (6.44$\pm$ 0.25\%)\\ \hline
\textsc{partial ring} &55 (11.29$\pm$ 0.34\%)&42 (8.62$\pm$ 0.29\%)&13 (8.67$\pm$ 0.29\%)&8 (5.33$\pm$ 0.23\%)&68 (10.68$\pm$ 0.33\%)&50 (7.85$\pm$ 0.28\%)\\
\hline
\hline
\textsc{TOTAL} & 302 (62.02$\pm$0.79\%) & 185 (37.98$\pm$0.62\% ) & 92 (60.59$\pm$0.78\%) & 58 (38.66$\pm$0.62\%) & 394 (61.87$\pm$0.79\%) & 243  (38.15$\pm$0.62\%  \\ 
\hline
\end{tabular}
}
\label{tab:TabBar}
\end{table*}

\section{Control samples}

To examine the impact of group environments on ringed galaxies and gain insights into their behavior and properties, we obtained a control sample (CS) for ringed galaxies in poor and rich groups, respectively, composed of galaxies without ringed structures.
Following the study of \cite{Perez2009}, who showed that in order to obtain a suitable control sample, the redshift, morphology, magnitude, and local density environment must match. 

We began by selecting galaxies without rings that resided in groups but exhibited parameters similar to those of the overall sample of galaxies with rings in groups.
To achieve this, we used a Monte Carlo algorithm to match the redshift and r$-$band absolute magnitude ($M_r$) distributions of ringed and non-ringed galaxies in groups.
We also selected galaxies without rings with  concentration indices, C, similar to ringed galaxies to obtain a comparable bulge-to-disk ratio. By doing so, any difference in the results will be driven by the presence of rings and not by discrepancies associated with galaxy morphology.

\begin{figure}[h!]
\centering
\includegraphics[width=1.0\linewidth]{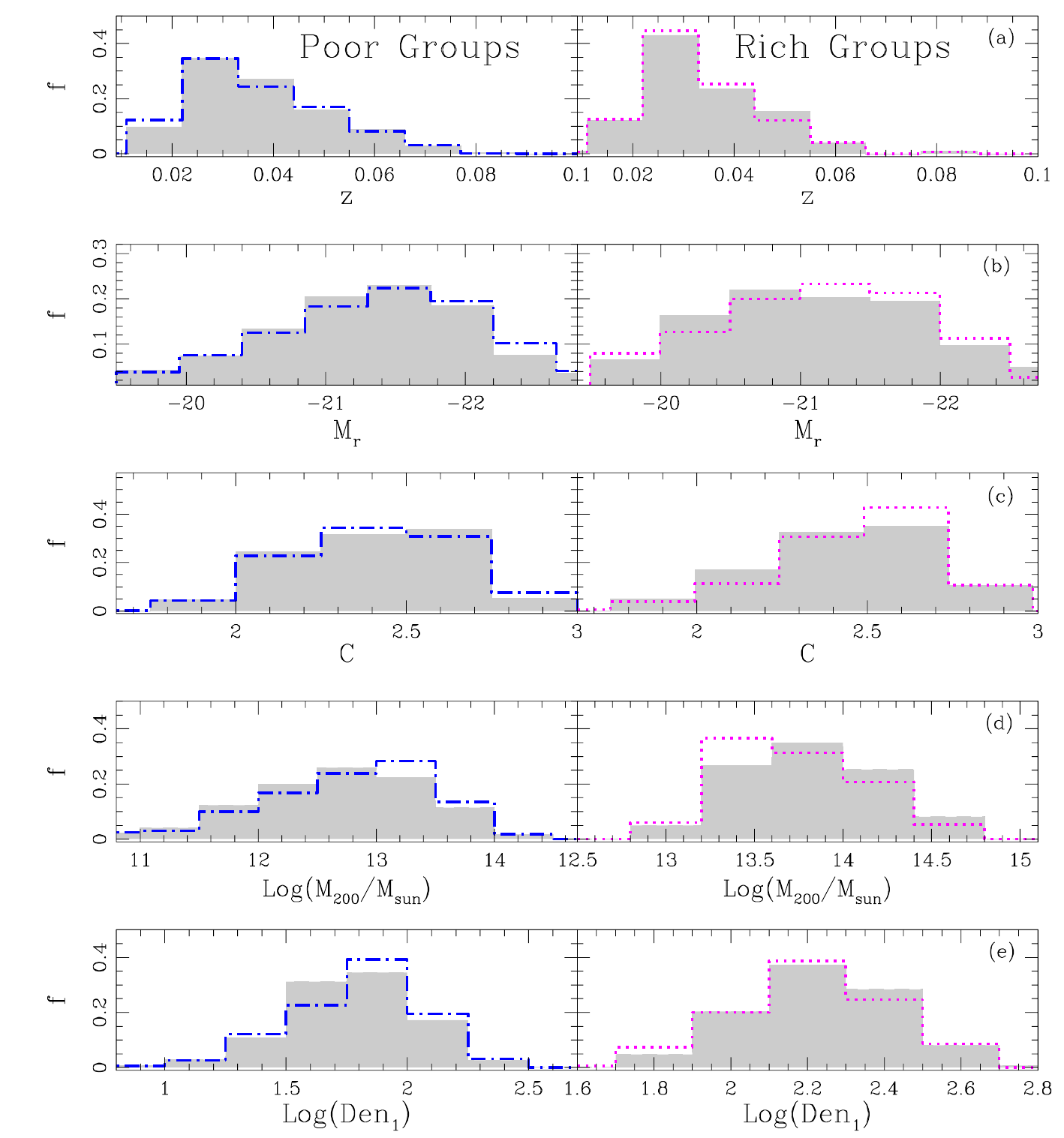}
\caption{Normalized distributions of (a) redshift, (b) ${r-}$ band absolute magnitude, (c) concentration parameter, (d) group mass, and (e) galaxy environmental density, for ringed galaxies in poor and rich groups (dot-dash blue and dotted magenta lines, respectively) and their corresponding control samples (shaded histograms).}
\label{control}
\end{figure}

In addition, we selected non-ringed galaxies in groups with a $Log (M_{200}/M_{\odot})$ distribution similar to that of the ringed ones. The group virial mass used for membership refinement by \cite{Tempel2017} was estimated by assuming a Navarro–Frenk–White (NFW) density profile, velocity dispersion, and group extent on the sky, given by {$M_{200} \propto \sigma_{v}^{2} \cdot \sigma_{sky}$}.

For the control sample, we also selected non-ringed galaxies with similar distribution of the density parameter as the ringed sample. To achieve this, we used the density parameter $Den_1$, calculated from the SDSS r-band luminosities, where different smoothing scales with varying lengths (1, 2, 4, and 8 $h^{-1}$ Mpc) indicate different environments (see Sect. 4 in \citealt{Tempel2012}). The smaller smoothing lengths represent group scales, while the larger smoothing lengths correspond to cluster and supercluster environments.
In this study, we adopted the normalized environmental density of ringed galaxies in the group, using a smoothing scale of $a=1h^{-1}$Mpc  as a local density estimator for the galaxies.
This further confirms that the local density environment is comparable for both ringed and non-ringed galaxies.

Subsequently, the obtained sample of galaxies without rings was divided into poor and rich groups, resulting in 514 (81\%) non-ringed galaxies in poor groups and 123 (19\%) in rich groups, as depicted in the lower columns of Table \ref{tab:Tabar}. These sub-samples serve as control samples for ringed galaxies in both poor and rich groups (see panels a, b, c, d, and e in Fig.~\ref{control}). In order to confirm this, we conducted the Kolmogorov-Smirnov (KS) test between ringed galaxies in rich and poor groups and their corresponding control samples. The obtained p-value was $p > 0 \period 05$, supporting the null hypothesis. Therefore, we rely that our study  will not be biased  by differences in the main properties of the ringed galaxies and the corresponding control samples.

Regarding the percentages of galaxies with rings in poor and rich groups, it has been observed that these are slightly different ($\sim$5$\sigma$ within the considered errors) from those in the control samples. Furthermore, while these findings suggest a trend toward a higher presence of ringed galaxies in poor groups, caution must be exercised in making definitive claims about the role the environment plays in their occurrence. Other potential factors related to the presence of rings in galaxies should also be taken into consideration.

\section{Bar frequency and distribution of ringed galaxies within groups}

An important aspect in the investigation of ringed galaxies is the relationship between the presence of rings and the existence of bars within these systems. Numerous studies have explored the connection between ringed galaxies and barred structures. \cite{Comeron2014} use mid-infrared observations to analyze the shape and orientation of rings and bars in galaxies from the $S^{4}G$ (Spitzer Survey of Stellar Structure in Galaxies). This comprehensive study led to the creation of ARRAKIS (Atlas of Resonance Rings as Known in the $S^{4}G$), an atlas focused on resonant rings in representative galaxies within the local Universe. Their statistical analysis  revealed that barred galaxies tend to exhibit a higher frequency of outer rings (1.7 times more) and inner rings (1.3 times more) when compared to non-barred galaxies. However, their results suggest that rings may exist independently of bars and that the possible mechanisms for ring formation in non-barred galaxies include the presence of weak ovals and long-lived spiral modes. Additionally, they concluded that it is plausible that some rings may be formed by bars that are no longer present. 

Several authors have also studied the role of the environment in the occurrence of bars \citep[e.g.,][]{Skibba2012,Corsini2013,Alonso2014}. Some studies have concluded that the environment does not play a significant role in the occurrence of bars, which are largely determined by the internal processes of the host galaxy, while others have inferred that bar frequency is likely influenced by the environment of the galaxy \citep{Aguerri2009,Li2009,Lee2012A,Skibba2012,Sarkar2021}. 

Since our sample of ringed galaxies was derived from the dataset compiled by F21, which  not only required us to identify rings, but also to conduct a visual examination of galaxies to determine the presence or absence of bars. Thus,  we decided to analyze the fraction of ringed galaxies with and without bars within rich and poor groups. From our sample we calculated numbers and percentages of barred ringed galaxies in group environments, finding that 302 ($\approx$ 62\%) of the ringed galaxies in poor groups are barred, and 185 ($\approx$ 38\%) do not present a barred structure, while in rich groups 92 ($\approx$ 61\%) are barred and 58 ($\approx$ 39\%) unbarred.
It is observed that $\approx$ 37\% of galaxies with inner rings present a bar. Moreover, about 13\% of galaxies featuring both outer and inner rings exhibit bars, while in galaxies with partial rings, bars are present in approximately 11\%. Remarkably, only 1\% and 0\% of the galaxies with outer and nuclear rings, respectively, display bar structures. We notice that for the different samples of ringed galaxies within groups, the percentages of barred galaxies are similar, regardless of whether the group is poor or rich (see Table \ref{tab:TabBar}).

In addition, we also studied the fraction of barred galaxies in the  control samples. The results revealed that $\approx$ 24\% of the control galaxies in rich groups and $\approx$ 19\% of the control galaxies in poor groups exhibit bars. When contrasting these findings with the fraction of observed bars in ringed galaxies, a clear relationship between the presence of rings and the existence of bars in galaxies becomes evident. In both rich and poor groups, barred galaxies are prevalent among ringed galaxies, while non-barred galaxies dominate among those without rings. These findings may be suggesting that the group environment is not the primary factor influencing the presence of bars in galaxies and that the formation of rings might be more closely related to internal processes within the galaxies, rather than being strongly influenced by the group environment \citep{Wilman2012,Buta2019}.

\section{\textbf{Ringed galaxy distribution in groups.}}

In a well-virialized and evolved cluster, the core presents an extreme density environment where particular physical mechanisms modify the morphology and galaxy evolution. In addition, the merging process tends to bring the most luminous galaxies closer to the central region \citep{Robotham2010}. 
As a result, several galaxies with signs of interactions are more concentrated towards the group centers \citep [e.g.,][]{elli2010, Alonso2004, Alonso2012}.
In this direction, the analysis of the fraction of ringed galaxies in denser regions of the group core, as compared to the outer zone of lower density, may provide clues about the relative effects of physical mechanisms driving the formation and evolution of galactic rings. 
Therefore, an interesting point is knowing if ringed galaxies have a particular location in poor and rich groups with respect to their corresponding control samples. 

Assuming a NFW profile, \cite{Tempel2017} estimated the group virial radius, $R_{200}$, as the radius of a sphere where the mean matter density is 200 times higher than the mean of the Universe. 
However, under the virialization assumption, some model dependent parameter, such as virial radius, are uncertain for groups with less than five members (\citealt{Li2019}, \citealt{Tempel2017}). 
To avoid bias in our analysis, we excluded groups with less than 5 members from our catalogs of poor groups, both in the ringed and control samples. After removing these groups, we were left with 191 poor groups containing ringed galaxies and 205 groups in the corresponding control sample. These groups were utilized in the analysis of properties, when it was deemed necessary. To address the impact of the removed groups on the distributions shown in Fig. \ref{control}, we conducted additional analyses to examine whether their exclusion influenced the overall results. Our analysis revealed that the exclusion of these galaxies did not significantly alter the overall patterns observed in the distributions.

We studied the distribution of the normalized projected distance to the group center, $d_{CG}/R_{200}$, for ringed galaxies in both rich and poor groups, along with the control samples. The results unveiled that ringed galaxies did not manifest a preferential location within the groups as compared to their non-ringed counterparts. Both categories exhibited analogous behavior, a pattern substantiated by the values of the K-S test, which indicated that the distributions remained comparable across all the considered scenarios.

%-----------------------------------------------------------------------------
\section{Ringed galaxy properties}

The aim of this section is to explore the effects of group environments on the features of host galaxies with ringed structures.
To accomplish this, we analyze several properties, such as the star formation activity, stellar populations, and colors of ringed galaxies. We compared these properties in poor and rich groups to suitable control samples that have been described previously.
 
\subsection{Star formation activity and stellar population}

\begin{figure*}
  \centering
  \includegraphics[width=0.95\textwidth]{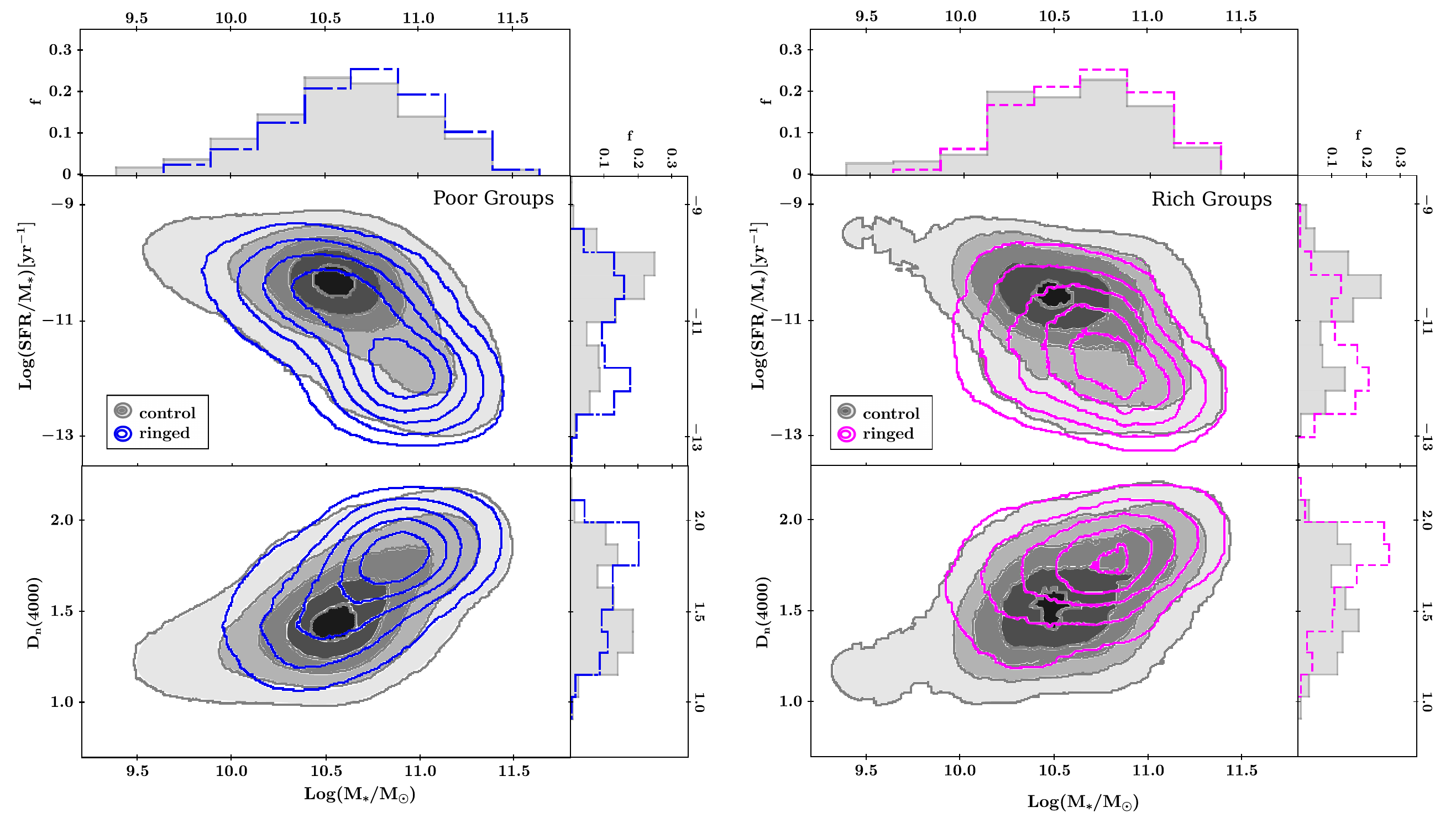}~\hfill
 \caption{$Log(SFR/M_*)$ and $D_n(4000)$ as a function of stellar masses. {Left panels:} Ringed galaxies in poor groups (represented by blue contours) and their corresponding control sample (represented by gray contours). The side panels show the normalized distributions for each sample. Top-most panel displays the normalized stellar mass distribution for the aforementioned samples.
{Right panels:} Same for ringed and non-ringed galaxies in rich groups (represented by magenta and gray contours, respectively).}
    \label{sfrM}
\end{figure*}

In order to assess the impact of environment on star formation and the stellar age populations of ringed galaxies inhabiting poor and rich groups, we used the specific star formation rate parameter,  $Log(SFR/M_*)$, as a suitable indicator of the star formation activity in our analysis.
This parameter is estimated as a function of the $H\alpha$ line luminosity and is normalized using stellar masses. The stellar masses included in this analysis were obtained by the  Max Planck Institute for Astrophysics/Johns Hopkins University (MPA/JHU) team through photometry fits and exhibit minor discrepancies compared to those derived by \cite{Kauffmann2003} and \cite{Gallazi2005} using spectral indices, with the differences being negligible. In addition, the star formation rates ($SFR$) were obtained from the MPA/JHU catalogue, which were calculated based on the methods of \cite{Brinchmann2004}. The $SFR$ estimation for star-forming galaxies follows the \cite{Charlot2001} model, while for other galaxy classes, such as AGN, composite, and low signal-to-noise (S/N) galaxies, we employed a similar methodology to that in \cite{Brinchmann2004}. Nevertheless, certain modifications have been implemented. Aperture corrections for SFRs are conducted by fitting stochastic models to the photometry of galaxies outer regions, thereby addressing biases observed in prior studies.  

We also incorporated the spectral index $D_{n}(4000)$ \citep{Kauffmann2003}, which estimates the age of stellar populations and is calculated from the spectral discontinuity occurring at 4000 $\AA $, resulting from an accumulation of numerous spectral lines in a narrow region of the spectrum that is particularly important in old stars.
In this analysis, we used the $D_n(4000)$ definition obtained by \cite{Balogh1999}, as the ratio of the average flux densities in the narrow continuum bands (3850-3950 $\r{A}$ and 4000-4100 $\r{A}$). This allows us to better determine the ages of stellar populations under investigation.

Figure \ref{sfrM} shows the $Log(SFR/M_*)$ and $D_n(4000)$ as a function of stellar masses of the  galaxies with ringed structures in poor and rich groups,  including the control samples. 
It can be observed that the distribution of ringed galaxies shows a trend towards lower star formation activities and higher stellar ages values. This trend is more significant for ringed galaxies inhabiting rich groups than those residing in poor groups. In contrast, the control samples of galaxies without ringed structures display a more scattered distribution, showing more efficient star formation activity and a younger stellar population. 
Additionally, the figure presents the normalized distributions of $Log(SFR/M_*)$ and $D_n(4000)$ for the different samples, which account for the observed behavior.
The difference between these distributions was also quantified by the Kolmogorov-Smirnov statistics (with a confidence of 99.8\%).
Furthermore, the stellar mass distributions for ringed and control samples in poor and rich groups are presented at the top of the figure, revealing similar trends, also supported by the KS test.

We can also observe that the value located near $Log(SFR/M_*)$ $\approx$ -10.6 divides the distributions into two populations.
Similarly, the galaxies in our samples show a bimodality in the stellar population around $D_n(4000)$ $\approx$ 1.5. 
Both of these values represent the average of the medians obtained from each of the control samples.
Table 4 quantifies the percentages of galaxies with low star-formation activity and an old stellar population in our samples, using these limits as reference. We considered the standard error, computed as the standard deviation of the sampling distribution of our variables.
This finding suggests that in group environments, ringed galaxies exhibit lower star formation activity and aging of the stellar populations compared to galaxies without rings in the control sample. This trend is more significant in rich groups.

\begin{table}
\centering
\caption{Percentages of ringed galaxies and the control sample (CS) in rich and poor groups with low star formation activity and old stellar populations, along with their corresponding standard errors.}
\begin{tabular}{|c c c| }
\hline
 Restrictions &  $Log(SFR/M_*)<-10 \period 6$ & $D_n(4000)>1 \period 5$ \\
\hline
\hline
\% Ringed in poor groups &  66.12$\pm$ 0.81\% & 70.23 $\pm$ 0.83\% \\
\% CS in poor groups & 43.39$\pm$ 0.66 \% & 46.69 $\pm$ 0.68\% \\
\hline
\% Ringed in rich groups & 80.66$\pm$ 0.90\% & 84.00 $\pm$ 0.92\% \\
\% CS in rich groups & 53.66$\pm$ 0.73 \% & 56.91 $\pm$ 0.75\% \\
\hline
\hline
\end{tabular}
{\small}
\label{tab:TabSFR}
\end{table}

In addition, Fig. \ref{sfrC} displays the relationship between $Log(SFR/M_*)$ and $D_n(4000)$ with respect to the $C$ parameter for both ringed and non-ringed galaxies in rich and poor groups. 
Errors were estimated using the Bootstrap method in both this figure and the subsequent ones. Each iteration of the Bootstrap creates a bootstrap sample by randomly selecting galaxies from the original data with a replacement and obtaining estimates for parameters, such as the mean and standard errors \citep{Barrow1984}.
We observed that earlier morphological types (higher $C$ values) exhibit low star formation efficiency and old stellar populations. Additionally, the figure reveals that the presence of ringed structures in galaxies leads to decreased star formation activity and aged stellar populations compared to galaxies without rings in the corresponding control samples. This effect is particularly noticeable in the galaxy samples associated with rich groups across all morphological types.

In order to provide a more comprehensive analysis of the different types of rings and their implications for the observed trends, we examined the normalized distributions of $Log(SFR/M_*)$ for galaxies displaying different ring types in both poor and rich groups in Fig. \ref{HSFR}. We compared these distributions with their respective control samples.
It is evident that, in general, galaxies with different ring types in group environments exhibit lower levels of star formation activity compared to galaxies without ringed structures. This trend is particularly pronounced for galaxies with outer, inner + outer, and partial rings in rich groups. In contrast, spiral galaxies with nuclear ring structures display values of $Log(SFR/M_*)$ similar to those of the control galaxies, indicating comparatively higher levels of star formation. This trend becomes more pronounced in poor groups.
Furthermore, in a similar vein, Fig. \ref{HDn} shows that ringed galaxies present older stellar populations compared to their control counterparts, irrespective of the ring type and the poor/rich group environment. Nevertheless, galaxies with inner and outer rings exhibit the highest $D_n(4000)$ values with a pronounced signal toward older stellar populations. On the other hand, galaxies hosting nuclear rings demonstrate a younger stellar population (lower $D_n(4000)$ values), similar to that of the control sample, especially in poor groups.
Moreover, in all panels of Figs. \ref{HSFR} and \ref{HDn}, we measured the disparities of $Log(SFR/M_*)$ and $D_n(4000)$  between the distributions of ringed galaxies with different types of rings and their respective control samples in the poor and rich groups. The significance level achieved for these comparisons is 99.99\%, as determined by the KS statistics. The resulting D and $p$-values from the KS test reveal statistically significant differences between the distributions for most cases.

The significant difference between nuclear rings and other classes of rings in terms of $SFR$ and $D_n(4000)$ mainly lies  in the fact that nuclear rings are usually sites of star formation \citep{Buta1996,Knapen2005,Comeron2010}, often dominating the entire star formation activity of their host galaxies. Furthermore, observations of nearby galaxies have demonstrated that the environments of nuclear rings can host large populations of young massive star clusters \citep{Mazzuca2008,Ma2018}. In addition, the $SFR$ in nuclear rings is affected by many factors, such as the strength of non-axisymmetric perturbations in galaxies, the inflow rate of gas, and the strength of the magnetic field \citep{Yang2022,Seo2013,Seo2019}. For example, star formation rates appear to be lower in galaxies with nuclear rings and strong magnetic fields, as in those with strong bars \citep{Tabatabaei2018,Ma2018}.

F21 demonstrated that the existence of a ringed structure was correlated with a reduction in star formation activity. In particular, our analysis of star formation in ringed galaxies located in group environments shows similar results to those found in F21, further reinforcing these trends. 
Consequently, the outcomes of this research reveal that, on a global scale, ringed galaxies in denser environments experience a decrease in their star formation rate. In denser environments, it is plausible that more massive galaxies tend to have lower star formation rates because they are more likely to deplete their gas reservoirs through frequent interactions with other galaxies. On the other hand, in less dense environments such as poor groups, more massive galaxies may maintain higher star formation rates for a longer period due to a lower frequency of interactions that deplete their gas. However, when analyzing galaxies with specific type of rings, we find that environment may affect in a different way each ring category, which could be influencing the star formation activity. For nuclear rings, it is possible that galaxies are undergoing certain internal processes that favor gas inflow toward the nucleus, thereby increasing their star formation rate. In contrast, galaxies with outer and partial rings may be experiencing gas-depleting processes, leading to a decrease in star formation. Although some studies have shown that the star formation rate in outer rings appears to be low \citep{Kostiuk2016, Katkov2022}, this effect seems to be further accentuated by the environment.

\begin{figure}
  \centering
 \includegraphics[width=.5\textwidth]{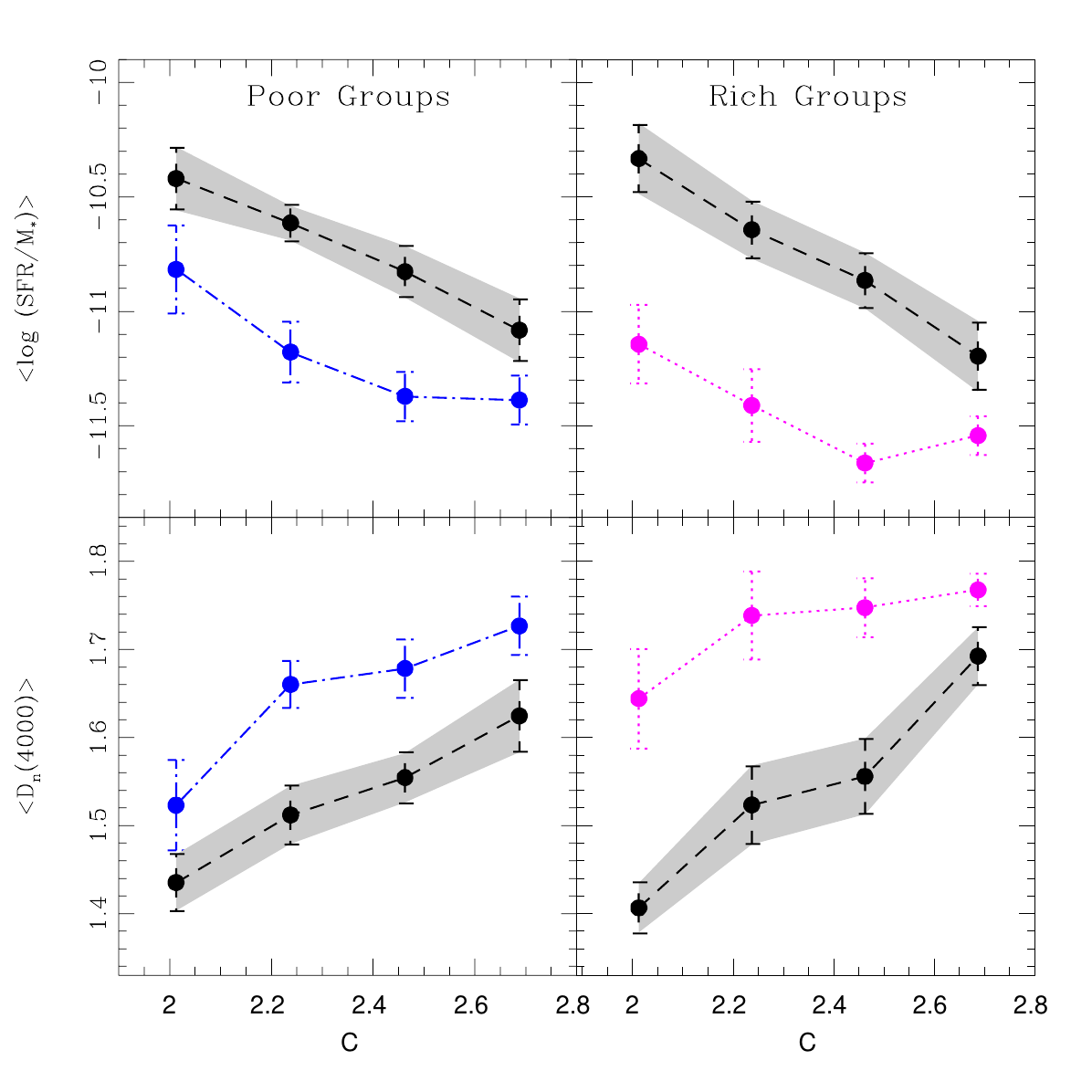}%\hfill 
\caption{ Mean values of  $Log(SFR/M_*)$ and $D_n(4000)$ as a function 
 of the concentration parameter, $C$, for ringed galaxies in poor groups (dot-dash blue lines), rich groups (dotted magenta lines), and their corresponding control samples (long-dash black lines).
 }
    \label{sfrC}
    \end{figure}

\begin{figure*}
  \centering
 \includegraphics[width=.90\textwidth]{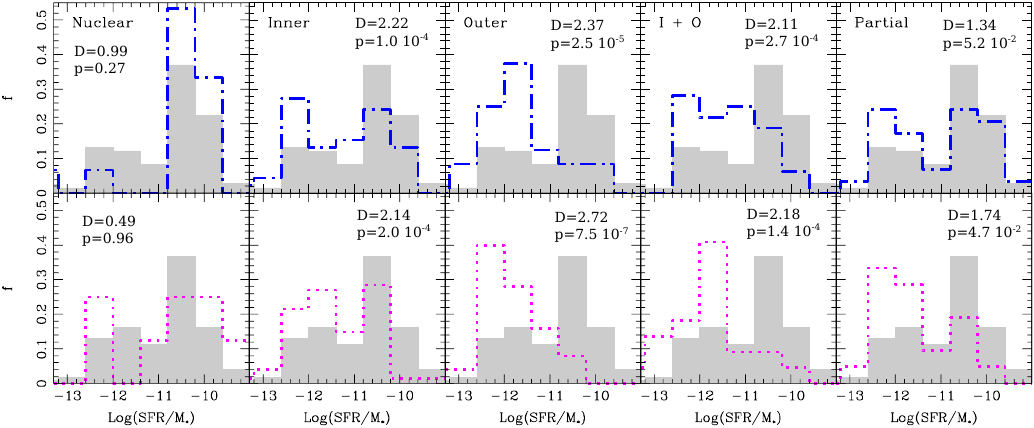}%\hfill 
\caption{$Log(SFR/M_*)$ normalized distributions of galaxies exhibiting different ring types (nuclear, inner, outer, inner+outer and partial) in poor and rich groups (upper and lower panels, respectively).} 
    \label{HSFR}
    \end{figure*}

\begin{figure*}
  \centering
 \includegraphics[width=.90\textwidth]{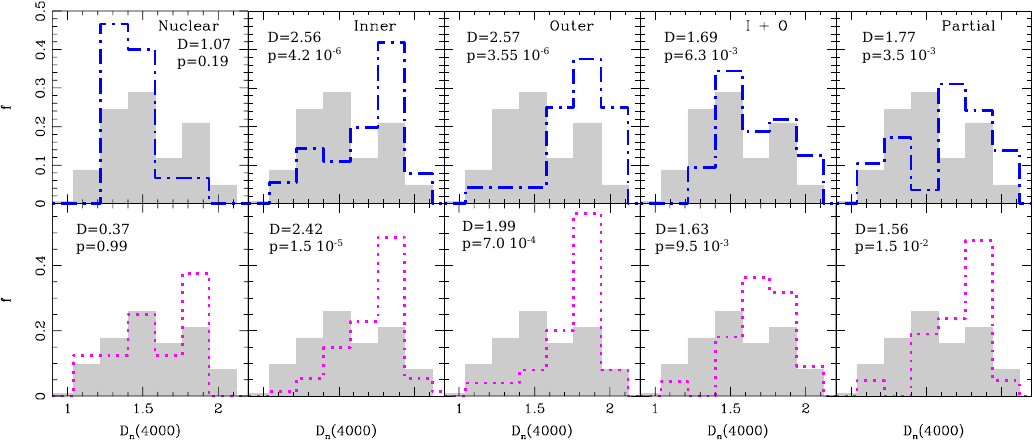}%\hfill 
\caption{$D_n(4000)$ normalized distributions of galaxies exhibiting different types of rings (nuclear, inner, outer, inner+outer and partial) in poor and rich groups (upper and lower panels, respectively).}

    \label{HDn}
    \end{figure*}

%-------------------------------------------------------------------------------------
\subsection{Galaxy colors}

\begin{figure*}
  \centering
  \includegraphics[width=.95\textwidth]{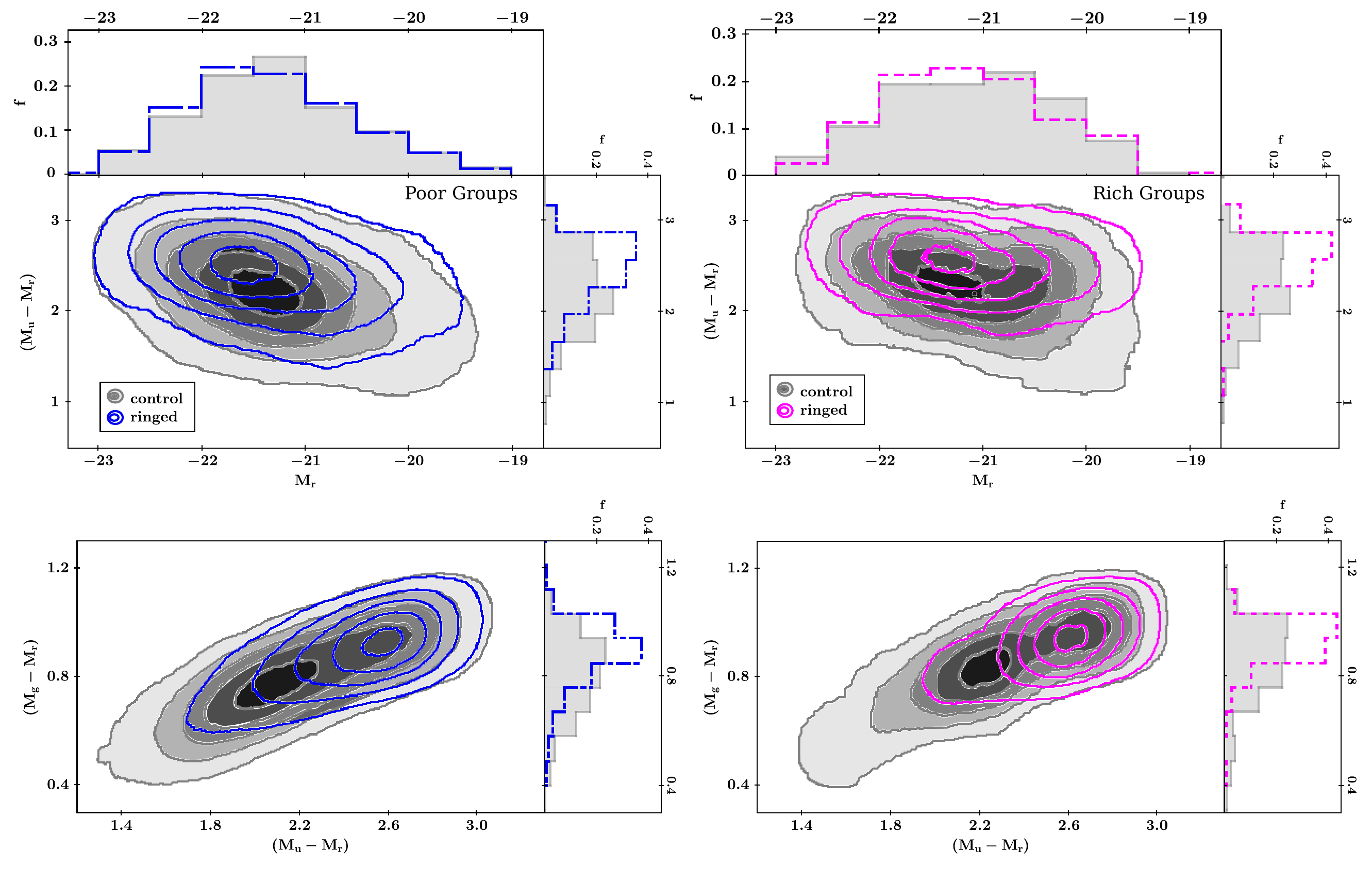}
 \caption{Color-magnitude and color-color diagrams for ringed galaxies in groups are displayed. 
  {Upper panels:} Color-magnitude diagrams for ringed galaxies in poor and rich groups (represented by blue and magenta contours, respectively), and their corresponding control samples (represented by gray contours). The vertical panels show the $M_u - M_r$  normalized distributions for each sample.
Top-most panels depict the normalized $M_r$ distributions for the same galaxy samples.
{Lower panels:} Color-color diagrams for ringed galaxies in poor groups (blue outlines), in rich groups (magenta outlines) and galaxies in their respective control samples (gray surface outlines). The vertical panels show the $M_g - M_r$ normalized distributions for each sample.}
\label{colM}
\end{figure*}

\begin{figure}
  \centering
 \includegraphics[width=.48\textwidth]{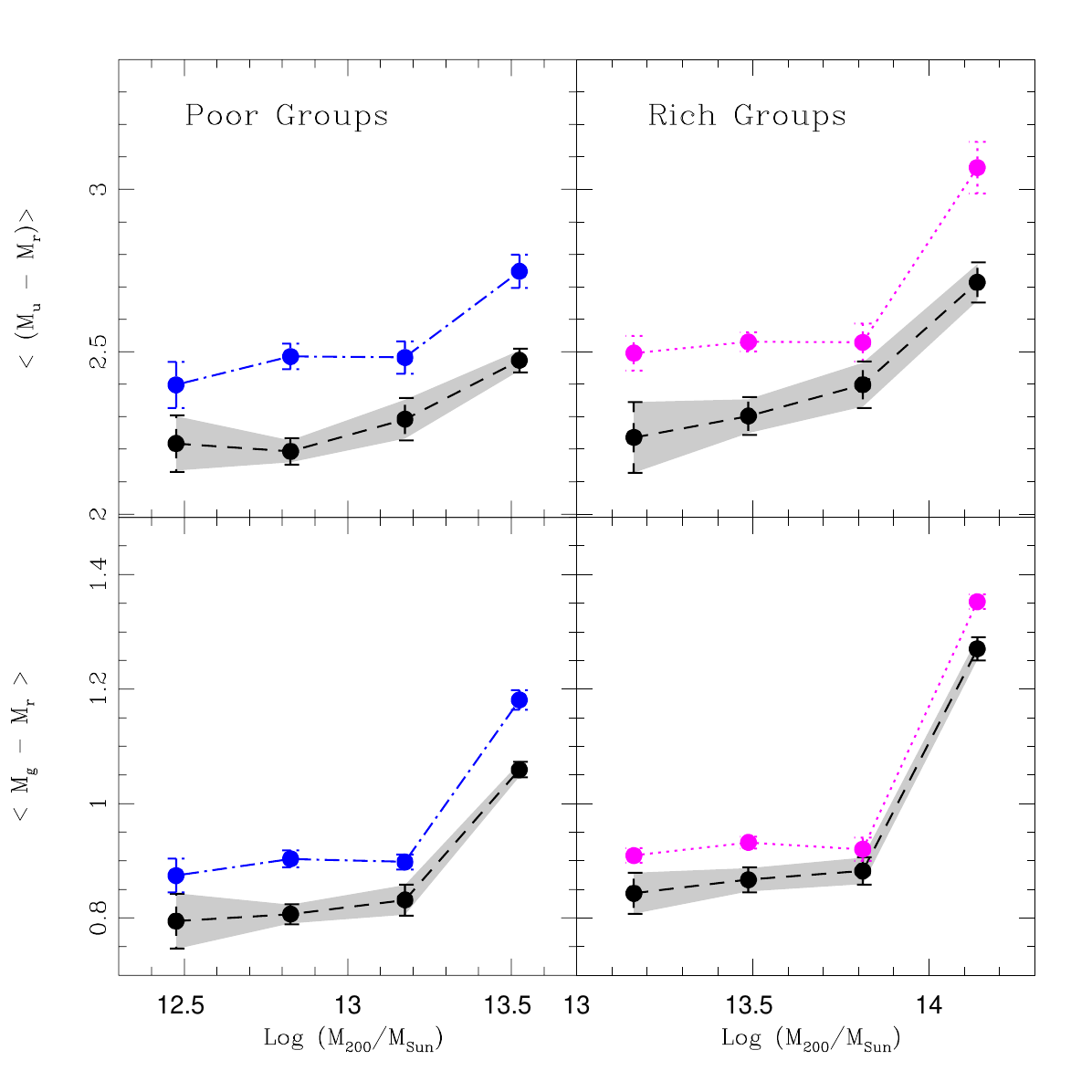}%\hfill 
\caption{Mean values of $(M_u - M_r)$ and $(M_g - M_r)$ as a function of the group masses, for ringed galaxies in poor groups (dot-dash blue lines), rich groups (dotted magenta lines) and their corresponding control samples (long-dash black lines).
}
    \label{ColMass}
    \end{figure}

\begin{figure*}
  \centering
 \includegraphics[width=.90\textwidth]{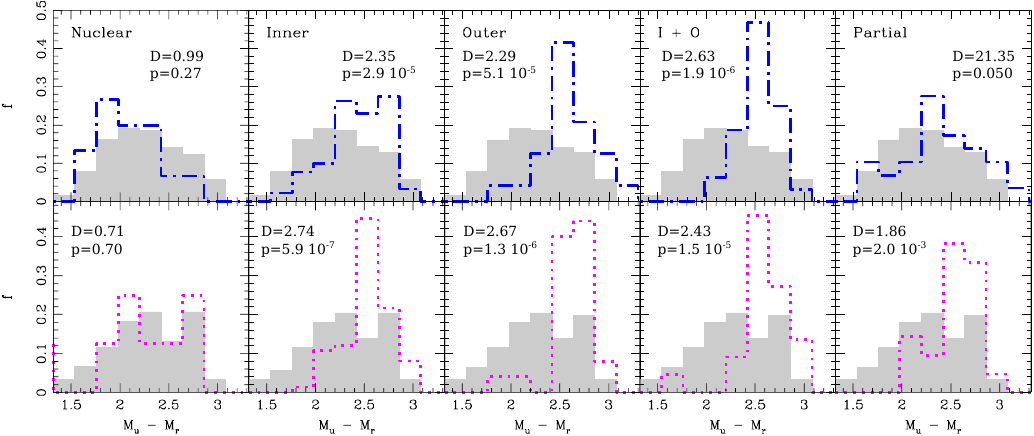}%\hfill 
\caption{$(M_u - M_r)$ normalized distributions of galaxies exhibiting different types of rings (nuclear, inner, outer, inner+outer and partial) in poor and rich groups (upper and lower panels, respectively).
}
    \label{HCol}
    \end{figure*}

Galaxy colors have a clear relation with the dominant stellar populations and morphology and, therefore, with star formation, gas content, and environment as well (\citealt{Feldmann2017,Pandey2020}).
In this context, a connection with the presence of ringed structures in disk galaxies is also plausible.
Then, in order to characterize the colors of ringed galaxies in group environments, in this section, we analyze the galaxy colors in both samples with and without rings that inhabit rich and poor groups.

In Fig. \ref {colM}, the color-magnitude diagram ($ M_u-M_r $ versus $M_r$) of these galaxies is shown, for ringed galaxies in rich groups and in poor groups. The corresponding control samples are also included. 
It can be observed that ringed galaxies are mostly found between the green valley  and the red sequence, with an excess of them in the red sequence, especially those in rich groups, while galaxies in the control samples tend to be located in the region of the green valley and the blue cloud.
The discrepancy between these distributions was also assessed using the Kolmogorov-Smirnov statistics, with a confidence level of 99.8\%.
In contrast, the $M_r$ distributions for both ringed and control samples within the poor and rich groups exhibit comparable patterns (see the top panels of Fig. \ref {colM}).
Moreover, we can observe that the value $ M_u-M_r $ $\approx$ 2.2 approximately separates the two peaks of each color distribution, in agreement with \cite{Strateva2001}.
This threshold can be considered to divide the blue and red galaxy populations. 
This limit is reflected in $ M_g-M_r $ $\approx$ 0.84, and also represents the median value of the distribution of galaxies in the control samples.
In this direction, with the aim of quantifying this tendency, we considered the excess of red color indexes ($ M_u-M_r > $ 2.2 and $ M_g-M_r > $ 0.84) of ringed galaxies in rich and poor groups with respect to the non-ringed objects in the respective control samples (see Table \ref{tab:Color}). In addition, the lower panels of the Fig. \ref {colM} show color-color diagrams for ringed galaxies in poor and rich groups and galaxies in the control samples. It is evident that there is a considerable surplus of ringed galaxies with red colors compared to non-ringed galaxies in the control samples. This trend is particularly pronounced among galaxies with ring structures inhabiting rich groups, which have redder colors, indicating older and less active populations. The trend is also present among ringed galaxies in poor groups, although it is less pronounced. Conversely, galaxies in the control samples show a slightly higher proportion of blue colors than the other samples.

Furthermore, Fig. \ref{ColMass} shows the correlation between the mean color indices, $M_u - M_r$ and $M_g - M_r$, relative to the host group masses for ringed galaxies found in both poor and rich groups. The results for the control samples are also displayed (represented by dashed lines). It is evident that the number of redder galaxies increases towards higher group masses. Furthermore, galaxies with ringed structures consistently exhibit redder colors compared to galaxies without rings in their respective control samples. Additionally, galaxies located within rich groups with $Log (M_{200}/M_{sun}$) $>$ 13.8 display highly reddened populations, and this trend is particularly evident among those with ringed structures.

In Fig. \ref{HCol}, we display a detailed analysis of the impact of the different ring types on the observed galaxy color trends, for galaxies in poor and rich group environments. We present the $M_u - M_r$ distributions of galaxies exhibiting different ring types in poor and rich groups, as compared with their respective control sample.
It is evident that galaxies with different ring types, located within poor and rich groups, display redder colors compared to galaxies without ringed structures. This trend is particularly significant in galaxies with inner, outer, and inner + outer rings in rich groups. However, spirals hosting nuclear ring structures in poor groups exhibit bluer populations similar to those of their respective control samples.
Additionally, the comparisons of the $M_u - M_r$ distributions from all panels of the Fig. \ref{HCol} between the ringed galaxies with different types of rings and their corresponding control samples have yielded a significance level of 99.99\% through the application of KS statistics.

In our previous study (F21), we reported that ringed galaxies exhibit significant differences in their properties compared to non-ringed ones. Particularly, we found that there is an excess of ringed galaxies with red colors and that these effects are more pronounced for ringed galaxies that have inner rings and bars with respect to their counterparts that have some other types of rings and are non-barred. Furthermore, the color-magnitude and color-color diagrams show that ringed galaxies are mostly concentrated in the red region, while non-ringed spiral objects are more extended to the blue zone. 
These results extend to ringed galaxies residing in  both rich and poor groups, suggesting that a ring may be a part of a process that contributes to the consumption of gas in the disk, leading to suppressed star formation activity and also resulting in the aging and reddening of the stellar population.
Furthermore, common processes found in higher density environments, such as ram pressure stripping, evaporation, and tidal interactions that remove gas from disks \citep{Hashimoto1998}, can accelerate gas consumption in ringed galaxies residing in groups. The accelerated consumption of gas in the rings of galaxies due to interactions with the environment can affect the color of galaxies by decreasing star formation, aging and reddening the existing stellar population.

Our results are slightly different from those of \citet{Kelvin2018}, who visually classified a sample of galaxies in the local universe ($z<0 \period 06$), with stellar masses in the range of $10 \period 25<Log(M_{*}/M_{sun})<10 \period 75$, and examined variations in galaxy structure depending on morphology and galaxy color. They found an excess of ringed galaxies in the green valley, compared to the red sequence and the blue cloud.
On the other hand, authors like \cite{Holwerda2022}  investigated the clustering of a galaxy sample by mapping the feature space of the Galaxy and Mass Assembly (GAMA) sample onto a self-organizing map. They adopted the same color criterion as \citet{Kelvin2018} to define the green valley (see \citealt{Kelvin2018,Bremer2018}, for further details), but considered a wider mass range. These authors found that the green valley is populated by several interstitial sub-populations of galaxies, consisting of both elliptical and spiral galaxies. Furthermore, the multiple populations of the green valley, interspersed between the red and blue populations on the self-organizing map, support the idea that the green valley galaxy population is an intermediate population that does not necessarily represent a single population transitioning from one space to another;  rather, it would occupy a niche between the main galaxy populations \citep{Holwerda2022}.
When the authors of \cite{Holwerda2022} constrained their analysis to the mass range used by \cite{Kelvin2018} and \cite{Bremer2018}, they realized that their sample was significantly reduced to red galaxies. In a similar way, we tested by restricting ourselves to this mass range and color cuts (see \cite{Kelvin2018}), which resulted in a significant decrease in galaxies, leaving us with almost no galaxies in the blue cloud. Considering this point, we believe that any differences between our results and those of \cite{Kelvin2018} could be primarily attributed to the different criteria used, such as mass ranges, color thresholds used to define each region, and the inclusion of elliptical galaxies in their statistics, which are not considered in this work.

\begin{table} 
\center
\caption{Percentages of ringed galaxies and the control sample (CS) in rich and poor groups with red colors, and their standard errors.}
\begin{tabular}{|c c c| }
\hline
Restrictions &  $ (M_u-M_r)>2\period2 $ & $ (M_g-M_r)>0\period 84 $ \\
\hline
\hline
\% Ringed in poor groups &  75.56$\pm$ 0.87  \% & 71.05$\pm$ 0.84\%   \\
\% CS in poor groups & 49.02$\pm$ 0.70\% & 43.19 $\pm$ 0.66\%  \\
\hline
\% Ringed in rich groups & 87.33$\pm$0.93 \% & 86.00$\pm$0.93\%  \\
\% CS in rich groups & 58.54 $\pm$ 0.77\% & 56.10 $\pm$ 0.75\%  \\
\hline
\hline
\end{tabular}
{\small}
\label{tab:Color}
\end{table}

%------------------------------------------------------------------------------------

\begin{figure}[!h]
\centering
\includegraphics[width=0.51\textwidth]{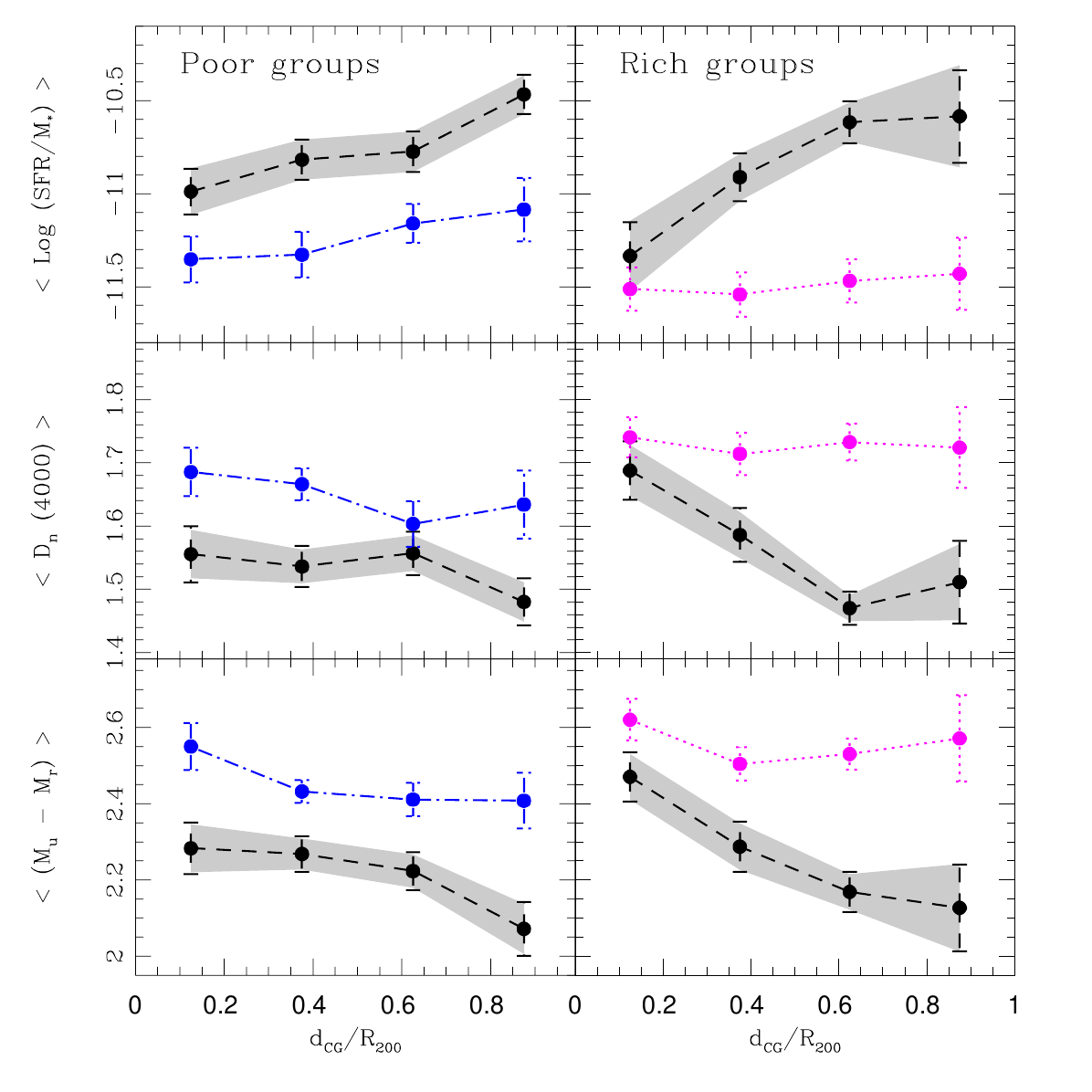}~\hfill
 \caption{Mean values of $Log(SFR/M_*)$, $D_n(4000)$ and $(M_u - M_r)$ as a function of the normalized group-centric distance, $d_{CG}/R_{200}$, in poor groups (dot-dash blue lines), rich groups (dotted magenta lines), and their respective control samples (long-dash black lines).
 }
\label{Prodc}
\end{figure}

\begin{figure}[!h]
\centering
 \includegraphics[width=0.51\textwidth]{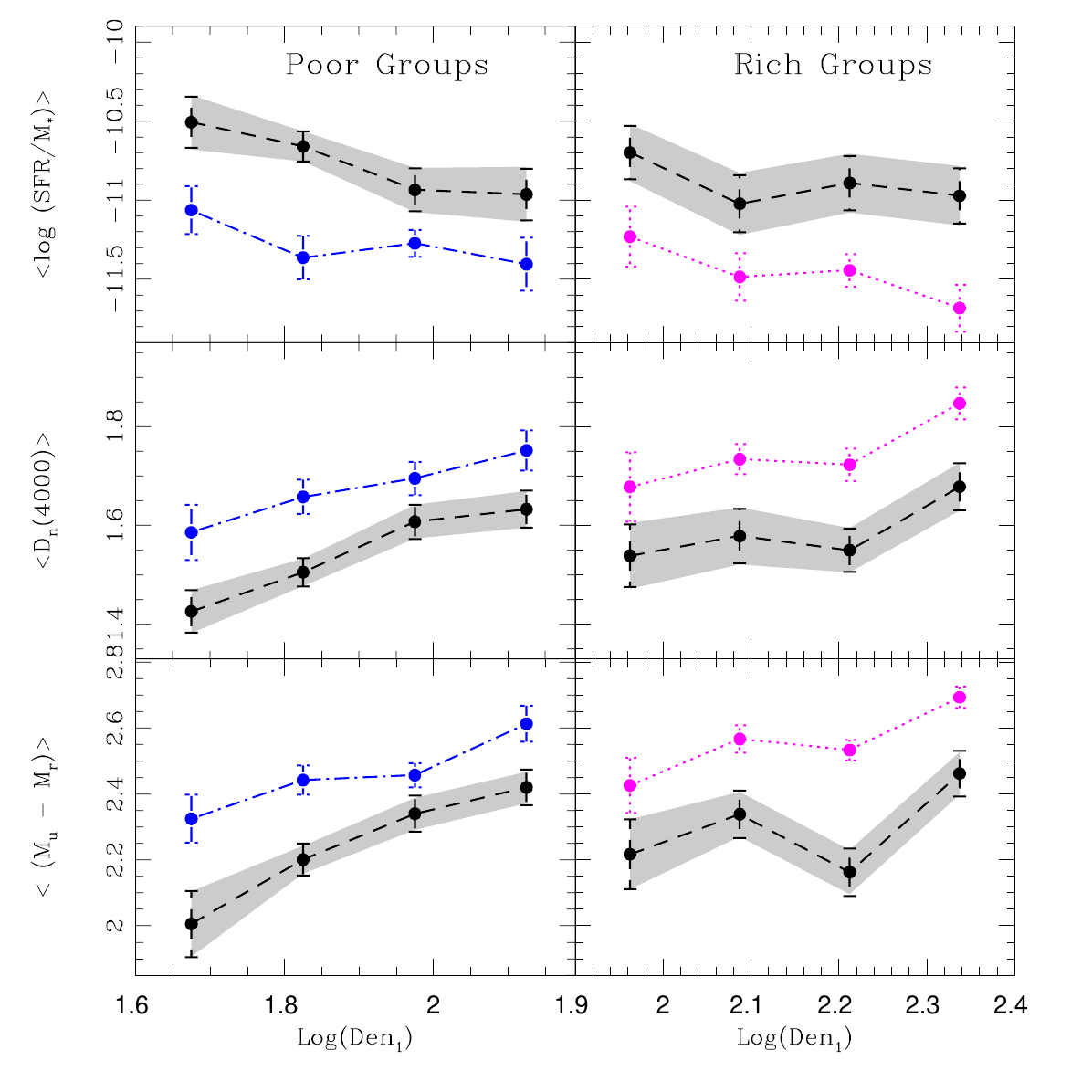}
 % ~\hfill
 \caption{Mean values of  $Log(SFR/M_*)$, $D_n(4000)$ and $(M_u - M_r)$ as a function of the galaxy environmental density in poor groups (dot-dash blue lines), rich groups (dotted magenta lines), and their respective  control samples (long-dash black lines).
 }
\label{ltot}
\end{figure}

\subsection{Dependence on the group-centric distance and environmental density}

In this section, we explore the relationship between star formation activity, color, and age of stellar population in relation to group-centric distance and environmental density, to enhance our understanding of how different features of ringed galaxies behave in group environments.
It has been well established that the central regions of galaxy groups and clusters are dominated by galaxies with redder colors compared to those in the outer regions \citep{Rodriguez2022A&A,Mercurio2021}.

Figure \ref{Prodc} displays the mean $Log(SFR/M_*)$, $D_n(4000)$ and $M_u - M_r$ plotted against the normalized group-centric distance, $d_{CG}/R_{200}$, for both ringed and non-ringed galaxies in poor and rich groups.
We can observe that ringed galaxies have lower star formation activity, older stellar populations, and redder colors compared to their non-ringed counterparts in the control samples across all distances from the group center. This trend is more evident in ringed galaxies located in poor groups; however, interestingly, in rich groups, their distribution over the entire range of distances to the group center appears relatively constant, suggesting that the variation of these properties are independent regardless of their position within rich groups.

In addition, we analyzed the mean values of $Log(SFR/M_*)$, $D_n(4000)$, and $M_u - M_r$ with respect to the environmental density ($Den_1$) of galaxies with rings in poor and rich groups, as well as their respective control samples (see Fig. \ref{ltot}).
It may be noticed that as the environmental density increases, galaxies exhibit a decrement in their star formation activity, along with an increment in the age of the stellar population and a shift towards redder colors.
Moreover, ringed galaxies in both poor and rich groups display less efficient star formation activity compared to galaxies without rings in their respective control samples, across all environmental density ranges. Additionally, across the entire $Den_1$ range, galaxies with ringed structures residing in poor and rich groups tend to present an older and more reddish stellar population compared to their corresponding control samples.

In this context, the lower star formation activity, older stellar populations, and redder colors observed in ringed galaxies compared to their non-ringed counterparts suggest that the presence of a ring structure may be indicative of a different formation history or evolutionary pathway. The observed trend of decreasing star formation activity, older stellar population ages and a shift towards redder colors with increasing environmental density is consistent with the idea that galaxy properties are affected by the surrounding environment.

\section{Summary and conclusions}

In this work, we analyzed the properties of ringed galaxies in dense environments corresponding to poor and rich groups of galaxies. 

We identified ringed galaxies, considering inner, outer, nuclear, inner+outer, and partial rings, in group density environments by cross-correlation of the ringed galaxy catalog from F21 with galaxies in the groups catalog from \cite{Tempel2017}, obtaining a sample of 637 ringed galaxies that reside in groups. This value represents a fraction of 34.1\% relative to the full sample of 1868 ringed galaxies. 
We divided the resulting sample into two subsamples based on the richness of the group: ringed galaxies in poor groups ($3  \leq N_{rich}  \leq 10$) and ringed galaxies in rich groups ($11  \leq N_{rich}  \leq 50$). Our analysis revealed that out of the total sample of 637 ringed galaxies residing in group environments, $\approx$ 76\%  were found in poor groups, whereas only about 24\% were present in rich systems.
We observed that galaxies with inner rings are the most common in both rich and poor groups, constituting $\approx$ 47\% of the sample. On the other hand, galaxies with nuclear rings were found to be the least frequent in both groups.

Another notable observation is that outer rings in our samples are more common in galaxies residing in rich groups, while partial rings decrease in these groups, suggesting that partial rings might originate primarily from internal processes within the galaxy and then be influenced by the presence of tidal forces.

To investigate the correlation between the  dense environment and the characteristics of ringed galaxies, we constructed suitable control samples of galaxies without ringed structures by simultaneously matching the redshift, r-band absolute magnitude, morphology, group masses, and environmental density with those of the ringed galaxies.
These control samples were used to reliably estimate differences between galaxies with and without rings, thus helping to reveal the role played by the group environments on ringed galaxy properties.
Regarding the non-ring galaxies in the control samples, 81\% were found in poor groups, and 19\% were in rich groups. Our results also revealed a stronger preference of ringed galaxies toward poor groups. While we must exercise caution in making definitive statements, the implication of this finding suggests that the environment might exert minimal influence on the occurrence of rings. Instead, internal factors could be more responsible for the formation and presence of these intriguing structures.

We also examined the occurrence of ringed galaxies with and without bars within rich and poor groups. The incidence of barred galaxies exhibited remarkable similarity regardless of the group richness. These findings suggest that the presence of bars in ringed galaxies appears to be independent of the surrounding environment. Moreover, based on the distribution of group-centric distance, we found that ringed galaxies did not show any difference in their distribution within the rich/poor groups compared to the corresponding control samples.

Furthermore, we explored the impact of ringed structures on the host galaxy properties residing in group environments. Our analysis revealed that galaxies possessing ringed structures display decreased levels of star formation activity and aging in their stellar populations when compared to their non-ringed counterparts. This trend is more pronounced for ringed galaxies within rich groups in comparison to those in poor ones. In contrast, galaxies without ringed structures in the control samples exhibit more efficient star formation activity and a younger stellar population. 
Although, in general, ringed galaxies exhibit a lower star formation rate compared to their non-ringed counterparts, our study reveals that galaxies with different types of rings respond differently to their environment. Galaxies possessing nuclear rings appear to undergo internal processes that promote the flow of gas towards the nucleus, enhancing star formation. Meanwhile, those with outer and partial rings are likely to experience gas-depleting mechanisms, leading to reduced rates of star formation.

An analysis of the color diagrams reveals that ringed galaxies predominantly populate the green valley and the red sequence, showing a surplus of galaxies in the red sequence, especially within rich groups. On the other hand, galaxies in the control samples tend to reside in the green valley region and the blue zone of the diagrams.
We also found that the fraction of galaxies with redder colors increases with higher host group masses, and across the entire group mass range, ringed galaxies consistently exhibit redder colors compared to non-ringed ones in their respective control samples.

Our studies also indicate that ringed galaxies, regardless of their location within the group, exhibit lower levels of star formation activity, along with older stellar populations and redder colors, as compared to non-ringed galaxies in the control samples.
From the correlation between ringed galaxy properties and the environmental density parameter, the results show that as density increases, galaxies have lower star formation activity, which is also reflected in older stellar populations and redder colors. This trend is particularly noticeable in ringed galaxies in both poor and rich groups.

Our examination of ringed galaxies in dense environments, including both rich and poor groups, has indicated that a ring could be a constituent of a process that leads to the depletion of gas in the disk. This, in turn, causes a reduction in star formation activity and contributes to the aging and reddening of the host stellar population.
Additionally, standard mechanisms typically observed in dense environments, such as ram pressure stripping, evaporation, and tidal interactions, which eliminate material from disks, can expedite the process of consuming gas in ringed galaxies residing in groups, resulting in accelerated star formation activity and galaxy evolution.

\section*{Acknowledgments}

This work was partially supported by the Consejo Nacional de Investigaciones Cient\'{\i}ficas y T\'ecnicas and the Secretar\'{\i}a de Ciencia y T\'ecnica de la Universidad Nacional de San Juan. V.M. also acknowledges  support  from  project  DIDULS Regular PR2353857.

Funding for the SDSS has been provided by the Alfred P. Sloan Foundation, the Participating Institutions, the National Science Foundation, the U.S. Department of Energy, the National Aeronautics and Space
Administration, the Japanese Monbukagakusho, the Max Planck Society, and the Higher Education Funding Council for England. The SDSS Web Site is http://www.sdss.org/.

The SDSS is managed by the Astrophysical Research Consortium for the Participating Institutions. The Participating Institutions are the American Museum of Natural History, Astrophysical Institute Potsdam, University of Basel, University of Cambridge, Case Western Reserve University, University of Chicago, Drexel University, Fermilab, the Institute for Advanced Study, the
Japan Participation Group, Johns Hopkins University, the Joint Institute for Nuclear Astrophysics, the Kavli Institute for Particle Astrophysics and Cosmology, the Korean Scientist Group, the Chinese Academy of Sciences (LAMOST), Los Alamos National Laboratory, the Max-Planck-Institute for Astronomy (MPIA), the Max-Planck-Institute for Astrophysics (MPA), New Mexico State University, Ohio State University, University of Pittsburgh, University of Portsmouth, Princeton University, the United States Naval Observatory, and the University of Washington.

\bibliographystyle{aa} % style aa.bst
\bibliography{biblio} % references Yourfile.bib

\begin{thebibliography}{88}
\expandafter\ifx\csname natexlab\endcsname\relax\def\natexlab#1{#1}\fi

\bibitem[{{Abolfathi} {et~al.}(2018){Abolfathi}, {Aguado}, {Aguilar}, {Allende
  Prieto}, {Almeida}, {Ananna}, {Anders}, {Anderson}, {Andrews}, {Anguiano},
  {Arag{\'o}n-Salamanca}, {Argudo-Fern{\'a}ndez}, {Armengaud}, {Ata},
  {Aubourg}, {Avila-Reese}, {Badenes}, {Bailey}, {Balland}, {Barger},
  {Barrera-Ballesteros}, {Bartosz}, {Bastien}, {Bates}, {Baumgarten},
  {Bautista}, {Beaton}, {Beers}, {Belfiore}, {Bender}, {Bernardi}, {Bershady},
  {Beutler}, {Bird}, {Bizyaev}, {Blanc}, {Blanton}, {Blomqvist}, {Bolton},
  {Boquien}, {Borissova}, {Bovy}, {Bradna Diaz}, {Brandt}, {Brinkmann},
  {Brownstein}, {Bundy}, {Burgasser}, {Burtin}, {Busca}, {Ca{\~n}as},
  {Cano-D{\'\i}az}, {Cappellari}, {Carrera}, {Casey}, {Cervantes Sodi}, {Chen},
  {Cherinka}, {Chiappini}, {Choi}, {Chojnowski}, {Chuang}, {Chung}, {Clerc},
  {Cohen}, {Comerford}, {Comparat}, {Correa do Nascimento}, {da Costa},
  {Cousinou}, {Covey}, {Crane}, {Cruz-Gonzalez}, {Cunha}, {da Silva Ilha},
  {Damke}, {Darling}, {Davidson}, {Dawson}, {de Icaza Lizaola}, {de la
  Macorra}, {de la Torre}, {De Lee}, {de Sainte Agathe}, {Deconto Machado},
  {Dell'Agli}, {Delubac}, {Diamond-Stanic}, {Donor}, {Downes}, {Drory}, {du Mas
  des Bourboux}, {Duckworth}, {Dwelly}, {Dyer}, {Ebelke}, {Davis Eigenbrot},
  {Eisenstein}, {Elsworth}, {Emsellem}, {Eracleous}, {Erfanianfar},
  {Escoffier}, {Fan}, {Fern{\'a}ndez Alvar}, {Fernandez-Trincado}, {Fernando
  Cirolini}, {Feuillet}, {Finoguenov}, {Fleming}, {Font-Ribera}, {Freischlad},
  {Frinchaboy}, {Fu}, {G{\'o}mez Maqueo Chew}, {Galbany}, {Garc{\'\i}a
  P{\'e}rez}, {Garcia-Dias}, {Garc{\'\i}a-Hern{\'a}ndez}, {Garma Oehmichen},
  {Gaulme}, {Gelfand}, {Gil-Mar{\'\i}n}, {Gillespie}, {Goddard}, {Gonz{\'a}lez
  Hern{\'a}ndez}, {Gonzalez-Perez}, {Grabowski}, {Green}, {Grier}, {Gueguen},
  {Guo}, {Guy}, {Hagen}, {Hall}, {Harding}, {Hasselquist}, {Hawley}, {Hayes},
  {Hearty}, {Hekker}, {Hernandez}, {Hernandez Toledo}, {Hogg},
  {Holley-Bockelmann}, {Holtzman}, {Hou}, {Hsieh}, {Hunt}, {Hutchinson},
  {Hwang}, {Jimenez Angel}, {Johnson}, {Jones}, {J{\"o}nsson}, {Jullo}, {Khan},
  {Kinemuchi}, {Kirkby}, {Kirkpatrick}, {Kitaura}, {Knapp}, {Kneib},
  {Kollmeier}, {Lacerna}, {Lane}, {Lang}, {Law}, {Le Goff}, {Lee}, {Li}, {Li},
  {Lian}, {Liang}, {Lima}, {Lin}, {Long}, {Lucatello}, {Lundgren}, {Mackereth},
  {MacLeod}, {Mahadevan}, {Maia}, {Majewski}, {Manchado}, {Maraston},
  {Mariappan}, {Marques-Chaves}, {Masseron}, {Masters}, {McDermid}, {McGreer},
  {Melendez}, {Meneses-Goytia}, {Merloni}, {Merrifield}, {Meszaros}, {Meza},
  {Minchev}, {Minniti}, {Mueller}, {Muller-Sanchez}, {Muna}, {Mu{\~n}oz},
  {Myers}, {Nair}, {Nandra}, {Ness}, {Newman}, {Nichol}, {Nidever},
  {Nitschelm}, {Noterdaeme}, {O'Connell}, {Oelkers}, {Oravetz}, {Oravetz},
  {Ort{\'\i}z}, {Osorio}, {Pace}, {Padilla}, {Palanque-Delabrouille},
  {Palicio}, {Pan}, {Pan}, {Parikh}, {P{\^a}ris}, {Park}, {Peirani},
  {Pellejero-Ibanez}, {Penny}, {Percival}, {Perez-Fournon}, {Petitjean},
  {Pieri}, {Pinsonneault}, {Pisani}, {Prada}, {Prakash}, {Queiroz}, {Raddick},
  {Raichoor}, {Barboza Rembold}, {Richstein}, {Riffel}, {Riffel}, {Rix},
  {Robin}, {Rodr{\'\i}guez Torres}, {Rom{\'a}n-Z{\'u}{\~n}iga}, {Ross},
  {Rossi}, {Ruan}, {Ruggeri}, {Ruiz}, {Salvato}, {S{\'a}nchez}, {S{\'a}nchez},
  {Sanchez Almeida}, {S{\'a}nchez-Gallego}, {Santana Rojas}, {Santiago},
  {Schiavon}, {Schimoia}, {Schlafly}, {Schlegel}, {Schneider}, {Schuster},
  {Schwope}, {Seo}, {Serenelli}, {Shen}, {Shen}, {Shetrone}, {Shull}, {Silva
  Aguirre}, {Simon}, {Skrutskie}, {Slosar}, {Smethurst}, {Smith}, {Sobeck},
  {Somers}, {Souter}, {Souto}, {Spindler}, {Stark}, {Stassun}, {Steinmetz},
  {Stello}, {Storchi-Bergmann}, {Streblyanska}, {Stringfellow}, {Su{\'a}rez},
  {Sun}, {Szigeti}, {Taghizadeh-Popp}, {Talbot}, {Tang}, {Tao}, {Tayar},
  {Tembe}, {Teske}, {Thakar}, {Thomas}, {Tissera}, {Tojeiro}, {Tremonti},
  {Troup}, {Urry}, {Valenzuela}, {van den Bosch}, {Vargas-Gonz{\'a}lez},
  {Vargas-Maga{\~n}a}, {Vazquez}, {Villanova}, {Vogt}, {Wake}, {Wang},
  {Weaver}, {Weijmans}, {Weinberg}, {Westfall}, {Whelan}, {Wilcots}, {Wild},
  {Williams}, {Wilson}, {Wood-Vasey}, {Wylezalek}, {Xiao}, {Yan}, {Yang},
  {Ybarra}, {Y{\`e}che}, {Zakamska}, {Zamora}, {Zarrouk}, {Zasowski}, {Zhang},
  {Zhao}, {Zhao}, {Zheng}, {Zheng}, {Zhou}, {Zhu}, {Zinn}, \&
  {Zou}}]{Abolfathi2018}
{Abolfathi}, B., {Aguado}, D.~S., {Aguilar}, G., {et~al.} 2018, \apjs, 235, 42

\bibitem[{{Aguerri} {et~al.}(2009){Aguerri}, {M{\'e}ndez-Abreu}, \&
  {Corsini}}]{Aguerri2009}
{Aguerri}, J.~A.~L., {M{\'e}ndez-Abreu}, J., \& {Corsini}, E.~M. 2009, \aap,
  495, 491

\bibitem[{{Alam} {et~al.}(2015){Alam}, {Albareti}, {Allende Prieto}, {Anders},
  {Anderson}, {Anderton}, {Andrews}, {Armengaud}, {Aubourg}, {Bailey}, {Basu},
  {Bautista}, {Beaton}, {Beers}, {Bender}, {Berlind}, {Beutler}, {Bhardwaj},
  {Bird}, {Bizyaev}, {Blake}, {Blanton}, {Blomqvist}, {Bochanski}, {Bolton},
  {Bovy}, {Shelden Bradley}, {Brandt}, {Brauer}, {Brinkmann}, {Brown},
  {Brownstein}, {Burden}, {Burtin}, {Busca}, {Cai}, {Capozzi}, {Carnero
  Rosell}, {Carr}, {Carrera}, {Chambers}, {Chaplin}, {Chen}, {Chiappini},
  {Chojnowski}, {Chuang}, {Clerc}, {Comparat}, {Covey}, {Croft}, {Cuesta},
  {Cunha}, {da Costa}, {Da Rio}, {Davenport}, {Dawson}, {De Lee}, {Delubac},
  {Deshpande}, {Dhital}, {Dutra-Ferreira}, {Dwelly}, {Ealet}, {Ebelke},
  {Edmondson}, {Eisenstein}, {Ellsworth}, {Elsworth}, {Epstein}, {Eracleous},
  {Escoffier}, {Esposito}, {Evans}, {Fan}, {Fern{\'a}ndez-Alvar}, {Feuillet},
  {Filiz Ak}, {Finley}, {Finoguenov}, {Flaherty}, {Fleming}, {Font-Ribera},
  {Foster}, {Frinchaboy}, {Galbraith-Frew}, {Garc{\'\i}a},
  {Garc{\'\i}a-Hern{\'a}ndez}, {Garc{\'\i}a P{\'e}rez}, {Gaulme}, {Ge},
  {G{\'e}nova-Santos}, {Georgakakis}, {Ghezzi}, {Gillespie}, {Girardi},
  {Goddard}, {Gontcho}, {Gonz{\'a}lez Hern{\'a}ndez}, {Grebel}, {Green},
  {Grieb}, {Grieves}, {Gunn}, {Guo}, {Harding}, {Hasselquist}, {Hawley},
  {Hayden}, {Hearty}, {Hekker}, {Ho}, {Hogg}, {Holley-Bockelmann}, {Holtzman},
  {Honscheid}, {Huber}, {Huehnerhoff}, {Ivans}, {Jiang}, {Johnson},
  {Kinemuchi}, {Kirkby}, {Kitaura}, {Klaene}, {Knapp}, {Kneib}, {Koenig},
  {Lam}, {Lan}, {Lang}, {Laurent}, {Le Goff}, {Leauthaud}, {Lee}, {Lee},
  {Licquia}, {Liu}, {Long}, {L{\'o}pez-Corredoira}, {Lorenzo-Oliveira},
  {Lucatello}, {Lundgren}, {Lupton}, {Mack}, {Mahadevan}, {Maia}, {Majewski},
  {Malanushenko}, {Malanushenko}, {Manchado}, {Manera}, {Mao}, {Maraston},
  {Marchwinski}, {Margala}, {Martell}, {Martig}, {Masters}, {Mathur},
  {McBride}, {McGehee}, {McGreer}, {McMahon}, {M{\'e}nard}, {Menzel},
  {Merloni}, {M{\'e}sz{\'a}ros}, {Miller}, {Miralda-Escud{\'e}}, {Miyatake},
  {Montero-Dorta}, {More}, {Morganson}, {Morice-Atkinson}, {Morrison},
  {Mosser}, {Muna}, {Myers}, {Nandra}, {Newman}, {Neyrinck}, {Nguyen},
  {Nichol}, {Nidever}, {Noterdaeme}, {Nuza}, {O'Connell}, {O'Connell},
  {O'Connell}, {Ogando}, {Olmstead}, {Oravetz}, {Oravetz}, {Osumi}, {Owen},
  {Padgett}, {Padmanabhan}, {Paegert}, {Palanque-Delabrouille}, {Pan},
  {Parejko}, {P{\^a}ris}, {Park}, {Pattarakijwanich}, {Pellejero-Ibanez},
  {Pepper}, {Percival}, {P{\'e}rez-Fournon}, {P{\'e}rez-R{\`a}fols},
  {Petitjean}, {Pieri}, {Pinsonneault}, {Porto de Mello}, {Prada}, {Prakash},
  {Price-Whelan}, {Protopapas}, {Raddick}, {Rahman}, {Reid}, {Rich}, {Rix},
  {Robin}, {Rockosi}, {Rodrigues}, {Rodr{\'\i}guez-Torres}, {Roe}, {Ross},
  {Ross}, {Rossi}, {Ruan}, {Rubi{\~n}o-Mart{\'\i}n}, {Rykoff},
  {Salazar-Albornoz}, {Salvato}, {Samushia}, {S{\'a}nchez}, {Santiago},
  {Sayres}, {Schiavon}, {Schlegel}, {Schmidt}, {Schneider}, {Schultheis},
  {Schwope}, {Sc{\'o}ccola}, {Scott}, {Sellgren}, {Seo}, {Serenelli}, {Shane},
  {Shen}, {Shetrone}, {Shu}, {Silva Aguirre}, {Sivarani}, {Skrutskie},
  {Slosar}, {Smith}, {Sobreira}, {Souto}, {Stassun}, {Steinmetz}, {Stello},
  {Strauss}, {Streblyanska}, {Suzuki}, {Swanson}, {Tan}, {Tayar}, {Terrien},
  {Thakar}, {Thomas}, {Thomas}, {Thompson}, {Tinker}, {Tojeiro}, {Troup},
  {Vargas-Maga{\~n}a}, {Vazquez}, {Verde}, {Viel}, {Vogt}, {Wake}, {Wang},
  {Weaver}, {Weinberg}, {Weiner}, {White}, {Wilson}, {Wisniewski},
  {Wood-Vasey}, {Ye`che}, {York}, {Zakamska}, {Zamora}, {Zasowski}, {Zehavi},
  {Zhao}, {Zheng}, {Zhou}, {Zhou}, {Zou}, \& {Zhu}}]{Alam2015}
{Alam}, S., {Albareti}, F.~D., {Allende Prieto}, C., {et~al.} 2015, \apjs, 219,
  12

\bibitem[{{Allington-Smith} {et~al.}(1993){Allington-Smith}, {Ellis}, {Zirbel},
  \& {Oemler}}]{Allington1993}
{Allington-Smith}, J.~R., {Ellis}, R., {Zirbel}, E.~L., \& {Oemler}, Augustus,
  J. 1993, \apj, 404, 521

\bibitem[{{Alonso} {et~al.}(2004){Alonso}, {Tissera}, {Coldwell}, \&
  {Lambas}}]{Alonso2004}
{Alonso}, M.~S., {Tissera}, P.~B., {Coldwell}, G., \& {Lambas}, D.~G. 2004,
  \mnras, 352, 1081

\bibitem[{{Alonso} {et~al.}(2014){Alonso}, {Coldwell}, \&
  {Lambas}}]{Alonso2014}
{Alonso}, S., {Coldwell}, G., \& {Lambas}, D.~G. 2014, \aap, 572, A86

\bibitem[{{Alonso} {et~al.}(2012){Alonso}, {Mesa}, {Padilla}, \&
  {Lambas}}]{Alonso2012}
{Alonso}, S., {Mesa}, V., {Padilla}, N., \& {Lambas}, D.~G. 2012, \aap, 539,
  A46

\bibitem[{{Baldry} {et~al.}(2006){Baldry}, {Balogh}, {Bower}, {Glazebrook},
  {Nichol}, {Bamford}, \& {Budavari}}]{Baldry2006}
{Baldry}, I.~K., {Balogh}, M.~L., {Bower}, R.~G., {et~al.} 2006, \mnras, 373,
  469

\bibitem[{{Balogh} {et~al.}(2004){Balogh}, {Baldry}, {Nichol}, {Miller},
  {Bower}, \& {Glazebrook}}]{Balogh_2004}
{Balogh}, M.~L., {Baldry}, I.~K., {Nichol}, R., {et~al.} 2004, \apjl, 615, L101

\bibitem[{{Balogh} {et~al.}(1999){Balogh}, {Morris}, {Yee}, {Carlberg}, \&
  {Ellingson}}]{Balogh1999}
{Balogh}, M.~L., {Morris}, S.~L., {Yee}, H.~K.~C., {Carlberg}, R.~G., \&
  {Ellingson}, E. 1999, \apj, 527, 54

\bibitem[{{Barrow} {et~al.}(1984){Barrow}, {Bhavsar}, \& {Sonoda}}]{Barrow1984}
{Barrow}, J.~D., {Bhavsar}, S.~P., \& {Sonoda}, D.~H. 1984, \mnras, 210, 19

\bibitem[{{Blanton} {et~al.}(2017){Blanton}, {Bershady}, {Abolfathi},
  {Albareti}, {Allende Prieto}, {Almeida}, {Alonso-Garc{\'\i}a}, {Anders},
  {Anderson}, {Andrews}, {Aquino-Ort{\'\i}z}, {Arag{\'o}n-Salamanca},
  {Argudo-Fern{\'a}ndez}, {Armengaud}, {Aubourg}, {Avila-Reese}, {Badenes},
  {Bailey}, {Barger}, {Barrera-Ballesteros}, {Bartosz}, {Bates}, {Baumgarten},
  {Bautista}, {Beaton}, {Beers}, {Belfiore}, {Bender}, {Berlind}, {Bernardi},
  {Beutler}, {Bird}, {Bizyaev}, {Blanc}, {Blomqvist}, {Bolton}, {Boquien},
  {Borissova}, {van den Bosch}, {Bovy}, {Brandt}, {Brinkmann}, {Brownstein},
  {Bundy}, {Burgasser}, {Burtin}, {Busca}, {Cappellari}, {Delgado Carigi},
  {Carlberg}, {Carnero Rosell}, {Carrera}, {Chanover}, {Cherinka}, {Cheung},
  {G{\'o}mez Maqueo Chew}, {Chiappini}, {Choi}, {Chojnowski}, {Chuang},
  {Chung}, {Cirolini}, {Clerc}, {Cohen}, {Comparat}, {da Costa}, {Cousinou},
  {Covey}, {Crane}, {Croft}, {Cruz-Gonzalez}, {Garrido Cuadra}, {Cunha},
  {Damke}, {Darling}, {Davies}, {Dawson}, {de la Macorra}, {Dell'Agli}, {De
  Lee}, {Delubac}, {Di Mille}, {Diamond-Stanic}, {Cano-D{\'\i}az}, {Donor},
  {Downes}, {Drory}, {du Mas des Bourboux}, {Duckworth}, {Dwelly}, {Dyer},
  {Ebelke}, {Eigenbrot}, {Eisenstein}, {Emsellem}, {Eracleous}, {Escoffier},
  {Evans}, {Fan}, {Fern{\'a}ndez-Alvar}, {Fernandez-Trincado}, {Feuillet},
  {Finoguenov}, {Fleming}, {Font-Ribera}, {Fredrickson}, {Freischlad},
  {Frinchaboy}, {Fuentes}, {Galbany}, {Garcia-Dias},
  {Garc{\'\i}a-Hern{\'a}ndez}, {Gaulme}, {Geisler}, {Gelfand},
  {Gil-Mar{\'\i}n}, {Gillespie}, {Goddard}, {Gonzalez-Perez}, {Grabowski},
  {Green}, {Grier}, {Gunn}, {Guo}, {Guy}, {Hagen}, {Hahn}, {Hall}, {Harding},
  {Hasselquist}, {Hawley}, {Hearty}, {Gonzalez Hern{\'a}ndez}, {Ho}, {Hogg},
  {Holley-Bockelmann}, {Holtzman}, {Holzer}, {Huehnerhoff}, {Hutchinson},
  {Hwang}, {Ibarra-Medel}, {da Silva Ilha}, {Ivans}, {Ivory}, {Jackson},
  {Jensen}, {Johnson}, {Jones}, {J{\"o}nsson}, {Jullo}, {Kamble}, {Kinemuchi},
  {Kirkby}, {Kitaura}, {Klaene}, {Knapp}, {Kneib}, {Kollmeier}, {Lacerna},
  {Lane}, {Lang}, {Law}, {Lazarz}, {Lee}, {Le Goff}, {Liang}, {Li}, {Li},
  {Lian}, {Lima}, {Lin}, {Lin}, {Bertran de Lis}, {Liu}, {de Icaza Lizaola},
  {Long}, {Lucatello}, {Lundgren}, {MacDonald}, {Deconto Machado}, {MacLeod},
  {Mahadevan}, {Geimba Maia}, {Maiolino}, {Majewski}, {Malanushenko},
  {Malanushenko}, {Manchado}, {Mao}, {Maraston}, {Marques-Chaves}, {Masseron},
  {Masters}, {McBride}, {McDermid}, {McGrath}, {McGreer}, {Medina Pe{\~n}a},
  {Melendez}, {Merloni}, {Merrifield}, {Meszaros}, {Meza}, {Minchev},
  {Minniti}, {Miyaji}, {More}, {Mulchaey}, {M{\"u}ller-S{\'a}nchez}, {Muna},
  {Munoz}, {Myers}, {Nair}, {Nandra}, {Correa do Nascimento}, {Negrete},
  {Ness}, {Newman}, {Nichol}, {Nidever}, {Nitschelm}, {Ntelis}, {O'Connell},
  {Oelkers}, {Oravetz}, {Oravetz}, {Pace}, {Padilla}, {Palanque-Delabrouille},
  {Alonso Palicio}, {Pan}, {Parejko}, {Parikh}, {P{\^a}ris}, {Park}, {Patten},
  {Peirani}, {Pellejero-Ibanez}, {Penny}, {Percival}, {Perez-Fournon},
  {Petitjean}, {Pieri}, {Pinsonneault}, {Pisani}, {Poleski}, {Prada},
  {Prakash}, {Queiroz}, {Raddick}, {Raichoor}, {Barboza Rembold}, {Richstein},
  {Riffel}, {Riffel}, {Rix}, {Robin}, {Rockosi}, {Rodr{\'\i}guez-Torres},
  {Roman-Lopes}, {Rom{\'a}n-Z{\'u}{\~n}iga}, {Rosado}, {Ross}, {Rossi}, {Ruan},
  {Ruggeri}, {Rykoff}, {Salazar-Albornoz}, {Salvato}, {S{\'a}nchez}, {Aguado},
  {S{\'a}nchez-Gallego}, {Santana}, {Santiago}, {Sayres}, {Schiavon}, {da Silva
  Schimoia}, {Schlafly}, {Schlegel}, {Schneider}, {Schultheis}, {Schuster},
  {Schwope}, {Seo}, {Shao}, {Shen}, {Shetrone}, {Shull}, {Simon}, {Skinner},
  {Skrutskie}, {Slosar}, {Smith}, {Sobeck}, {Sobreira}, {Somers}, {Souto},
  {Stark}, {Stassun}, {Stauffer}, {Steinmetz}, {Storchi-Bergmann},
  {Streblyanska}, {Stringfellow}, {Su{\'a}rez}, {Sun}, {Suzuki}, {Szigeti},
  {Taghizadeh-Popp}, {Tang}, {Tao}, {Tayar}, {Tembe}, {Teske}, {Thakar},
  {Thomas}, {Thompson}, {Tinker}, {Tissera}, {Tojeiro}, {Hernandez Toledo}, {de
  la Torre}, {Tremonti}, {Troup}, {Valenzuela}, {Martinez Valpuesta},
  {Vargas-Gonz{\'a}lez}, {Vargas-Maga{\~n}a}, {Vazquez}, {Villanova}, {Vivek},
  {Vogt}, {Wake}, {Walterbos}, {Wang}, {Weaver}, {Weijmans}, {Weinberg},
  {Westfall}, {Whelan}, {Wild}, {Wilson}, {Wood-Vasey}, {Wylezalek}, {Xiao},
  {Yan}, {Yang}, {Ybarra}, {Y{\`e}che}, {Zakamska}, {Zamora}, {Zarrouk},
  {Zasowski}, {Zhang}, {Zhao}, {Zheng}, {Zheng}, {Zhou}, {Zhou}, {Zhu},
  {Zoccali}, \& {Zou}}]{sdssiv}
{Blanton}, M.~R., {Bershady}, M.~A., {Abolfathi}, B., {et~al.} 2017, \aj, 154,
  28

\bibitem[{{Blanton} {et~al.}(2005){Blanton}, {Eisenstein}, {Hogg}, {Schlegel},
  \& {Brinkmann}}]{Blanton2005}
{Blanton}, M.~R., {Eisenstein}, D., {Hogg}, D.~W., {Schlegel}, D.~J., \&
  {Brinkmann}, J. 2005, \apj, 629, 143

\bibitem[{{Blanton} \& {Roweis}(2007)}]{Blanton2007}
{Blanton}, M.~R. \& {Roweis}, S. 2007, \aj, 133, 734

\bibitem[{{Botzler} {et~al.}(2004){Botzler}, {Snigula}, {Bender}, \&
  {Hopp}}]{Botzler2004}
{Botzler}, C.~S., {Snigula}, J., {Bender}, R., \& {Hopp}, U. 2004, \mnras, 349,
  425

\bibitem[{{Bremer} {et~al.}(2018){Bremer}, {Phillipps}, {Kelvin}, {De Propris},
  {Kennedy}, {Moffett}, {Bamford}, {Davies}, {Driver}, {H{\"a}u{\ss}ler},
  {Holwerda}, {Hopkins}, {James}, {Liske}, {Percival}, \&
  {Taylor}}]{Bremer2018}
{Bremer}, M.~N., {Phillipps}, S., {Kelvin}, L.~S., {et~al.} 2018, \mnras, 476,
  12

\bibitem[{{Brinchmann} {et~al.}(2004){Brinchmann}, {Charlot}, {White},
  {Tremonti}, {Kauffmann}, {Heckman}, \& {Brinkmann}}]{Brinchmann2004}
{Brinchmann}, J., {Charlot}, S., {White}, S.~D.~M., {et~al.} 2004, \mnras, 351,
  1151

\bibitem[{{Buta} \& {Combes}(1996)}]{Buta1996}
{Buta}, R. \& {Combes}, F. 1996, \fcp, 17, 95

\bibitem[{{Buta}(2017)}]{Buta2017}
{Buta}, R.~J. 2017, \mnras, 471, 4027

\bibitem[{{Buta} {et~al.}(2019){Buta}, {Verdes-Montenegro}, {Damas-Segovia},
  {Jones}, {Blasco}, {Fern{\'a}ndez-Lorenzo}, {Sanchez}, {Garrido},
  {Ramirez-Moreta}, \& {Sulentic}}]{Buta2019}
{Buta}, R.~J., {Verdes-Montenegro}, L., {Damas-Segovia}, A., {et~al.} 2019,
  \mnras, 488, 2175

\bibitem[{{Charlot} \& {Longhetti}(2001)}]{Charlot2001}
{Charlot}, S. \& {Longhetti}, M. 2001, \mnras, 323, 887

\bibitem[{{Comer{\'o}n} {et~al.}(2010){Comer{\'o}n}, {Knapen}, {Beckman},
  {Laurikainen}, {Salo}, {Mart{\'\i}nez-Valpuesta}, \& {Buta}}]{Comeron2010}
{Comer{\'o}n}, S., {Knapen}, J.~H., {Beckman}, J.~E., {et~al.} 2010, \mnras,
  402, 2462

\bibitem[{{Comer{\'o}n} {et~al.}(2014){Comer{\'o}n}, {Salo}, {Laurikainen},
  {Knapen}, {Buta}, {Herrera-Endoqui}, {Laine}, {Holwerda}, {Sheth}, {Regan},
  {Hinz}, {Mu{\~n}oz-Mateos}, {Gil de Paz}, {Men{\'e}ndez-Delmestre},
  {Seibert}, {Mizusawa}, {Kim}, {Erroz-Ferrer}, {Gadotti}, {Athanassoula},
  {Bosma}, \& {Ho}}]{Comeron2014}
{Comer{\'o}n}, S., {Salo}, H., {Laurikainen}, E., {et~al.} 2014, \aap, 562,
  A121

\bibitem[{{Corsini} {et~al.}(2013){Corsini}, {M{\'e}ndez-Abreu},
  {S{\'a}nchez-Janssen}, {Aguerri}, \& {Zarattini}}]{Corsini2013}
{Corsini}, E.~M., {M{\'e}ndez-Abreu}, J., {S{\'a}nchez-Janssen}, R., {Aguerri},
  J.~A.~L., \& {Zarattini}, S. 2013, Memorie della Societa Astronomica Italiana
  Supplementi, 25, 74

\bibitem[{{de Vaucouleurs}(1959)}]{deVaucouleurs1959}
{de Vaucouleurs}, G. 1959, Handbuch der Physik, 53, 275

\bibitem[{{de Vaucouleurs} {et~al.}(1976){de Vaucouleurs}, {de Vaucouleurs}, \&
  {Corwin}}]{deVaucouleurs1976}
{de Vaucouleurs}, G., {de Vaucouleurs}, A., \& {Corwin}, J.~R. 1976, Second
  reference catalogue of bright galaxies, 1976, 0

\bibitem[{{D{\'\i}az-Garc{\'\i}a} {et~al.}(2019){D{\'\i}az-Garc{\'\i}a},
  {D{\'\i}az-Su{\'a}rez}, {Knapen}, \& {Salo}}]{Diaz2019}
{D{\'\i}az-Garc{\'\i}a}, S., {D{\'\i}az-Su{\'a}rez}, S., {Knapen}, J.~H., \&
  {Salo}, H. 2019, \aap, 625, A146

\bibitem[{{D'Onghia} {et~al.}(2008){D'Onghia}, {Mapelli}, \&
  {Moore}}]{Donghia2008}
{D'Onghia}, E., {Mapelli}, M., \& {Moore}, B. 2008, \mnras, 389, 1275

\bibitem[{{Dressler}(1980)}]{Dressler1980}
{Dressler}, A. 1980, \apj, 236, 351

\bibitem[{{Einasto} \& {Einasto}(1987)}]{Einasto1987}
{Einasto}, M. \& {Einasto}, J. 1987, \mnras, 226, 543

\bibitem[{{Eisenstein} {et~al.}(2011){Eisenstein}, {Weinberg}, {Agol},
  {Aihara}, {Allende Prieto}, {Anderson}, {Arns}, {Aubourg}, {Bailey},
  {Balbinot}, {Barkhouser}, {Beers}, {Berlind}, {Bickerton}, {Bizyaev},
  {Blanton}, {Bochanski}, {Bolton}, {Bosman}, {Bovy}, {Brandt}, {Breslauer},
  {Brewington}, {Brinkmann}, {Brown}, {Brownstein}, {Burger}, {Busca},
  {Campbell}, {Cargile}, {Carithers}, {Carlberg}, {Carr}, {Chang}, {Chen},
  {Chiappini}, {Comparat}, {Connolly}, {Cortes}, {Croft}, {Cunha}, {da Costa},
  {Davenport}, {Dawson}, {De Lee}, {Porto de Mello}, {de Simoni}, {Dean},
  {Dhital}, {Ealet}, {Ebelke}, {Edmondson}, {Eiting}, {Escoffier}, {Esposito},
  {Evans}, {Fan}, {Femen{\'\i}a Castell{\'a}}, {Dutra Ferreira}, {Fitzgerald},
  {Fleming}, {Font-Ribera}, {Ford}, {Frinchaboy}, {Garc{\'\i}a P{\'e}rez},
  {Gaudi}, {Ge}, {Ghezzi}, {Gillespie}, {Gilmore}, {Girardi}, {Gott}, {Gould},
  {Grebel}, {Gunn}, {Hamilton}, {Harding}, {Harris}, {Hawley}, {Hearty},
  {Hennawi}, {Gonz{\'a}lez Hern{\'a}ndez}, {Ho}, {Hogg}, {Holtzman},
  {Honscheid}, {Inada}, {Ivans}, {Jiang}, {Jiang}, {Johnson}, {Jordan},
  {Jordan}, {Kauffmann}, {Kazin}, {Kirkby}, {Klaene}, {Knapp}, {Kneib},
  {Kochanek}, {Koesterke}, {Kollmeier}, {Kron}, {Lampeitl}, {Lang}, {Lawler},
  {Le Goff}, {Lee}, {Lee}, {Leisenring}, {Lin}, {Liu}, {Long}, {Loomis},
  {Lucatello}, {Lundgren}, {Lupton}, {Ma}, {Ma}, {MacDonald}, {Mack},
  {Mahadevan}, {Maia}, {Majewski}, {Makler}, {Malanushenko}, {Malanushenko},
  {Mandelbaum}, {Maraston}, {Margala}, {Maseman}, {Masters}, {McBride},
  {McDonald}, {McGreer}, {McMahon}, {Mena Requejo}, {M{\'e}nard},
  {Miralda-Escud{\'e}}, {Morrison}, {Mullally}, {Muna}, {Murayama}, {Myers},
  {Naugle}, {Neto}, {Nguyen}, {Nichol}, {Nidever}, {O'Connell}, {Ogando},
  {Olmstead}, {Oravetz}, {Padmanabhan}, {Paegert}, {Palanque-Delabrouille},
  {Pan}, {Pandey}, {Parejko}, {P{\^a}ris}, {Pellegrini}, {Pepper}, {Percival},
  {Petitjean}, {Pfaffenberger}, {Pforr}, {Phleps}, {Pichon}, {Pieri}, {Prada},
  {Price-Whelan}, {Raddick}, {Ramos}, {Reid}, {Reyle}, {Rich}, {Richards},
  {Rieke}, {Rieke}, {Rix}, {Robin}, {Rocha-Pinto}, {Rockosi}, {Roe},
  {Rollinde}, {Ross}, {Ross}, {Rossetto}, {S{\'a}nchez}, {Santiago}, {Sayres},
  {Schiavon}, {Schlegel}, {Schlesinger}, {Schmidt}, {Schneider}, {Sellgren},
  {Shelden}, {Sheldon}, {Shetrone}, {Shu}, {Silverman}, {Simmerer}, {Simmons},
  {Sivarani}, {Skrutskie}, {Slosar}, {Smee}, {Smith}, {Snedden}, {Stassun},
  {Steele}, {Steinmetz}, {Stockett}, {Stollberg}, {Strauss}, {Szalay},
  {Tanaka}, {Thakar}, {Thomas}, {Tinker}, {Tofflemire}, {Tojeiro}, {Tremonti},
  {Vargas Maga{\~n}a}, {Verde}, {Vogt}, {Wake}, {Wan}, {Wang}, {Weaver},
  {White}, {White}, {Wilson}, {Wisniewski}, {Wood-Vasey}, {Yanny}, {Yasuda},
  {Y{\`e}che}, {York}, {Young}, {Zasowski}, {Zehavi}, \&
  {Zhao}}]{Eisenstein2011}
{Eisenstein}, D.~J., {Weinberg}, D.~H., {Agol}, E., {et~al.} 2011, \aj, 142, 72

\bibitem[{{Elagali} {et~al.}(2018){Elagali}, {Lagos}, {Wong}, {Staveley-Smith},
  {Trayford}, {Schaller}, {Yuan}, \& {Abadi}}]{Elagali2018}
{Elagali}, A., {Lagos}, C. D.~P., {Wong}, O.~I., {et~al.} 2018, \mnras, 481,
  2951

\bibitem[{{Ellison} {et~al.}(2010){Ellison}, {Patton}, {Simard}, {McConnachie},
  {Baldry}, \& {Mendel}}]{elli2010}
{Ellison}, S.~L., {Patton}, D.~R., {Simard}, L., {et~al.} 2010, \mnras, 407,
  1514

\bibitem[{{Elmegreen} {et~al.}(1992){Elmegreen}, {Elmegreen}, {Combes}, \&
  {Bellin}}]{Elmegreen1992}
{Elmegreen}, D.~M., {Elmegreen}, B.~G., {Combes}, F., \& {Bellin}, A.~D. 1992,
  \aap, 257, 17

\bibitem[{{Feldmann} {et~al.}(2017){Feldmann}, {Quataert}, {Hopkins},
  {Faucher-Gigu{\`e}re}, \& {Kere{\v{s}}}}]{Feldmann2017}
{Feldmann}, R., {Quataert}, E., {Hopkins}, P.~F., {Faucher-Gigu{\`e}re}, C.-A.,
  \& {Kere{\v{s}}}, D. 2017, \mnras, 470, 1050

\bibitem[{{Fernandez} {et~al.}(2021){Fernandez}, {Alonso}, {Mesa}, {Duplancic},
  \& {Coldwell}}]{Fernandez2021}
{Fernandez}, J., {Alonso}, S., {Mesa}, V., {Duplancic}, F., \& {Coldwell}, G.
  2021, \aap, 653, A71

\bibitem[{{Few} \& {Madore}(1986)}]{Few1986}
{Few}, J. M.~A. \& {Madore}, B.~F. 1986, \mnras, 222, 673

\bibitem[{{Gallazzi} {et~al.}(2005){Gallazzi}, {Charlot}, {Brinchmann},
  {White}, \& {Tremonti}}]{Gallazi2005}
{Gallazzi}, A., {Charlot}, S., {Brinchmann}, J., {White}, S. D.~M., \&
  {Tremonti}, C.~A. 2005, \mnras, 362, 41

\bibitem[{{Gozaliasl} {et~al.}(2018){Gozaliasl}, {Finoguenov}, {Khosroshahi},
  {Henriques}, {Tanaka}, {Ilbert}, {Wuyts}, {McCracken}, \&
  {Montanari}}]{Gozaliasl2018}
{Gozaliasl}, G., {Finoguenov}, A., {Khosroshahi}, H.~G., {et~al.} 2018, \mnras,
  475, 2787

\bibitem[{{Grouchy} {et~al.}(2010){Grouchy}, {Buta}, {Salo}, \&
  {Laurikainen}}]{Grouchy2010}
{Grouchy}, R.~D., {Buta}, R.~J., {Salo}, H., \& {Laurikainen}, E. 2010, \aj,
  139, 2465

\bibitem[{{Gusev} \& {Park}(2003)}]{Gusev2003}
{Gusev}, A.~S. \& {Park}, M.~G. 2003, \aap, 410, 117

\bibitem[{{Hashimoto} {et~al.}(1998){Hashimoto}, {Oemler}, {Lin}, \&
  {Tucker}}]{Hashimoto1998}
{Hashimoto}, Y., {Oemler}, Augustus, J., {Lin}, H., \& {Tucker}, D.~L. 1998,
  \apj, 499, 589

\bibitem[{{Helmi} {et~al.}(2003){Helmi}, {Navarro}, {Meza}, {Steinmetz}, \&
  {Eke}}]{Helmi2003}
{Helmi}, A., {Navarro}, J.~F., {Meza}, A., {Steinmetz}, M., \& {Eke}, V.~R.
  2003, \apjl, 592, L25

\bibitem[{{Holwerda} {et~al.}(2022){Holwerda}, {Smith}, {Porter}, {Henry},
  {Porter-Temple}, {Cook}, {Pimbblet}, {Hopkins}, {Bilicki}, {Turner},
  {Acquaviva}, {Wang}, {Wright}, {Kelvin}, \& {Grootes}}]{Holwerda2022}
{Holwerda}, B.~W., {Smith}, D., {Porter}, L., {et~al.} 2022, \mnras, 513, 1972

\bibitem[{{Huchra} \& {Geller}(1982)}]{Huchra1982}
{Huchra}, J.~P. \& {Geller}, M.~J. 1982, \apj, 257, 423

\bibitem[{{Katkov} {et~al.}(2022){Katkov}, {Kniazev}, {Sil'chenko}, \&
  {Gasymov}}]{Katkov2022}
{Katkov}, I.~Y., {Kniazev}, A.~Y., {Sil'chenko}, O.~K., \& {Gasymov}, D. 2022,
  \aap, 658, A154

\bibitem[{{Kauffmann} {et~al.}(2003){Kauffmann}, {Heckman}, {White}, {Charlot},
  {Tremonti}, {Brinchmann}, {Bruzual}, {Peng}, {Seibert}, {Bernardi},
  {Blanton}, {Brinkmann}, {Castander}, {Cs{\'a}bai}, {Fukugita}, {Ivezic},
  {Munn}, {Nichol}, {Padmanabhan}, {Thakar}, {Weinberg}, \&
  {York}}]{Kauffmann2003}
{Kauffmann}, G., {Heckman}, T.~M., {White}, S. D.~M., {et~al.} 2003, \mnras,
  341, 33

\bibitem[{{Kauffmann} {et~al.}(2004){Kauffmann}, {White}, {Heckman},
  {M{\'e}nard}, {Brinchmann}, {Charlot}, {Tremonti}, \&
  {Brinkmann}}]{Kauffmann2004}
{Kauffmann}, G., {White}, S. D.~M., {Heckman}, T.~M., {et~al.} 2004, \mnras,
  353, 713

\bibitem[{{Kelvin} {et~al.}(2018){Kelvin}, {Bremer}, {Phillipps}, {James},
  {Davies}, {De Propris}, {Moffett}, {Percival}, {Baldry}, {Collins},
  {Alpaslan}, {Bland-Hawthorn}, {Brough}, {Cluver}, {Driver}, {Hashemizadeh},
  {Holwerda}, {Laine}, {Lara-Lopez}, {Liske}, {Maciejewski}, {Napolitano},
  {Penny}, {Popescu}, {Sansom}, {Sutherland}, {Taylor}, {van Kampen}, \&
  {Wang}}]{Kelvin2018}
{Kelvin}, L.~S., {Bremer}, M.~N., {Phillipps}, S., {et~al.} 2018, \mnras, 477,
  4116

\bibitem[{{Knapen}(2005)}]{Knapen2005}
{Knapen}, J.~H. 2005, \aap, 429, 141

\bibitem[{{Kormendy}(1979)}]{Kormedy1979}
{Kormendy}, J. 1979, \apj, 227, 714

\bibitem[{{Kostiuk} \& {Sil'chenko}(2016)}]{Kostiuk2016}
{Kostiuk}, I.~P. \& {Sil'chenko}, O.~K. 2016, Baltic Astronomy, 25, 331

\bibitem[{{Lacerna} {et~al.}(2022){Lacerna}, {Rodriguez}, {Montero-Dorta},
  {O'Mill}, {Cora}, {Artale}, {Ruiz}, {Hough}, \&
  {Vega-Mart{\'\i}nez}}]{Lacerna2022}
{Lacerna}, I., {Rodriguez}, F., {Montero-Dorta}, A.~D., {et~al.} 2022, \mnras,
  513, 2271

\bibitem[{{Lamb} {et~al.}(1993){Lamb}, {Gerber}, \& {Balsara}}]{Lamb1993}
{Lamb}, S.~A., {Gerber}, R.~A., \& {Balsara}, D.~S. 1993, in Evolution of
  Galaxies and their Environment, ed. J.~M. {Shull} \& H.~A. {Thronson},
  225--226

\bibitem[{{Lee} {et~al.}(2012){Lee}, {Park}, {Lee}, \& {Choi}}]{Lee2012A}
{Lee}, G.-H., {Park}, C., {Lee}, M.~G., \& {Choi}, Y.-Y. 2012, \apj, 745, 125

\bibitem[{{Li} {et~al.}(2009){Li}, {Gadotti}, {Mao}, \& {Kauffmann}}]{Li2009}
{Li}, C., {Gadotti}, D.~A., {Mao}, S., \& {Kauffmann}, G. 2009, \mnras, 397,
  726

\bibitem[{{Li} {et~al.}(2019){Li}, {Gu}, {Yuan}, {Bao}, {He}, \&
  {Bian}}]{Li2019}
{Li}, F., {Gu}, Y.-Z., {Yuan}, Q.-R., {et~al.} 2019, \mnras, 484, 3806

\bibitem[{{Lietzen} {et~al.}(2012){Lietzen}, {Tempel}, {Hein{\"a}m{\"a}ki},
  {Nurmi}, {Einasto}, \& {Saar}}]{Lietzen2012}
{Lietzen}, H., {Tempel}, E., {Hein{\"a}m{\"a}ki}, P., {et~al.} 2012, \aap

\bibitem[{{Lynds} \& {Toomre}(1976)}]{Lynds1976}
{Lynds}, R. \& {Toomre}, A. 1976, \apj, 209, 382

\bibitem[{{Ma} {et~al.}(2018){Ma}, {de Grijs}, \& {Ho}}]{Ma2018}
{Ma}, C., {de Grijs}, R., \& {Ho}, L.~C. 2018, \apj, 857, 116

\bibitem[{{Madore}(1980)}]{Madore1980}
{Madore}, B.~F. 1980, \aj, 85, 507

\bibitem[{{Mazzuca} {et~al.}(2008){Mazzuca}, {Knapen}, {Veilleux}, \&
  {Regan}}]{Mazzuca2008}
{Mazzuca}, L.~M., {Knapen}, J.~H., {Veilleux}, S., \& {Regan}, M.~W. 2008,
  \apjs, 174, 337

\bibitem[{{Mercurio} {et~al.}(2021){Mercurio}, {Rosati}, {Biviano},
  {Annunziatella}, {Girardi}, {Sartoris}, {Nonino}, {Brescia}, {Riccio},
  {Grillo}, {Balestra}, {Caminha}, {De Lucia}, {Gobat}, {Seitz}, {Tozzi},
  {Scodeggio}, {Vanzella}, {Angora}, {Bergamini}, {Borgani}, {Demarco},
  {Meneghetti}, {Strazzullo}, {Tortorelli}, {Umetsu}, {Fritz}, {Gruen},
  {Kelson}, {Lombardi}, {Maier}, {Postman}, {Rodighiero}, \&
  {Ziegler}}]{Mercurio2021}
{Mercurio}, A., {Rosati}, P., {Biviano}, A., {et~al.} 2021, \aap, 656, A147

\bibitem[{{Oemler}(1974)}]{Oemler1974}
{Oemler}, Augustus, J. 1974, \apj, 194, 1

\bibitem[{{Pandey} \& {Sarkar}(2020)}]{Pandey2020}
{Pandey}, B. \& {Sarkar}, S. 2020, \mnras, 498, 6069

\bibitem[{{Perez} {et~al.}(2009){Perez}, {Tissera}, \& {Blaizot}}]{Perez2009}
{Perez}, J., {Tissera}, P., \& {Blaizot}, J. 2009, \mnras, 397, 748

\bibitem[{{Robotham} {et~al.}(2010){Robotham}, {Phillipps}, \& {de
  Propris}}]{Robotham2010}
{Robotham}, A., {Phillipps}, S., \& {de Propris}, R. 2010, \mnras, 403, 1812

\bibitem[{{Rodr{\'\i}guez-Mart{\'\i}n}
  {et~al.}(2022){Rodr{\'\i}guez-Mart{\'\i}n}, {Gonz{\'a}lez Delgado},
  {Mart{\'\i}nez-Solaeche}, {D{\'\i}az-Garc{\'\i}a}, {de Amorim},
  {Garc{\'\i}a-Benito}, {P{\'e}rez}, {Cid Fernandes}, {Carrasco}, {Maturi},
  {Finoguenov}, {Lopes}, {Cortesi}, {Lucatelli}, {Diego}, {Chies-Santos},
  {Dupke}, {Jim{\'e}nez-Teja}, {V{\'\i}lchez}, {Abramo}, {Alcaniz},
  {Ben{\'\i}tez}, {Bonoli}, {Cenarro}, {Crist{\'o}bal-Hornillos}, {Ederoclite},
  {Hern{\'a}n-Caballero}, {L{\'o}pez-Sanjuan}, {Mar{\'\i}n-Franch}, {Mendes de
  Oliveira}, {Moles}, {Sodr{\'e}}, {Taylor}, {Varela}, {V{\'a}zquez Rami{\'o}},
  \& {M{\'a}rquez}}]{Rodriguez2022A&A}
{Rodr{\'\i}guez-Mart{\'\i}n}, J.~E., {Gonz{\'a}lez Delgado}, R.~M.,
  {Mart{\'\i}nez-Solaeche}, G., {et~al.} 2022, \aap, 666, A160

\bibitem[{{Sarkar} {et~al.}(2021){Sarkar}, {Pandey}, \&
  {Bhattacharjee}}]{Sarkar2021}
{Sarkar}, S., {Pandey}, B., \& {Bhattacharjee}, S. 2021, \mnras, 501, 994

\bibitem[{{Schweizer} {et~al.}(1987){Schweizer}, {Ford}, {Jedrzejewski}, \&
  {Giovanelli}}]{Schweizer1987}
{Schweizer}, F., {Ford}, W.~Kent, J., {Jedrzejewski}, R., \& {Giovanelli}, R.
  1987, \apj, 320, 454

\bibitem[{{Seo} \& {Kim}(2013)}]{Seo2013}
{Seo}, W.-Y. \& {Kim}, W.-T. 2013, \apj, 769, 100

\bibitem[{{Seo} {et~al.}(2019){Seo}, {Kim}, {Kwak}, {Hsieh}, {Han}, \&
  {Hopkins}}]{Seo2019}
{Seo}, W.-Y., {Kim}, W.-T., {Kwak}, S., {et~al.} 2019, \apj, 872, 5

\bibitem[{{Sil'chenko} {et~al.}(2018){Sil'chenko}, {Kostiuk}, {Burenkov}, \&
  {Parul}}]{Silchenko2018}
{Sil'chenko}, O., {Kostiuk}, I., {Burenkov}, A., \& {Parul}, H. 2018, \aap,
  620, L7

\bibitem[{{Skibba} {et~al.}(2012){Skibba}, {Masters}, {Nichol}, {Zehavi},
  {Hoyle}, {Edmondson}, {Bamford}, {Cardamone}, {Keel}, {Lintott}, \&
  {Schawinski}}]{Skibba2012}
{Skibba}, R.~A., {Masters}, K.~L., {Nichol}, R.~C., {et~al.} 2012, \mnras, 423,
  1485

\bibitem[{{Skibba} \& {Sheth}(2009)}]{Skibba2009}
{Skibba}, R.~A. \& {Sheth}, R.~K. 2009, \mnras, 392, 1080

\bibitem[{{Smirnov} \& {Reshetnikov}(2022)}]{smir22}
{Smirnov}, D.~V. \& {Reshetnikov}, V.~P. 2022, \mnras, 516, 3692

\bibitem[{{Smith} {et~al.}(2022){Smith}, {Calder{\'o}n-Castillo}, {Shin},
  {Raouf}, \& {Ko}}]{Smith2022}
{Smith}, R., {Calder{\'o}n-Castillo}, P., {Shin}, J., {Raouf}, M., \& {Ko}, J.
  2022, \aj, 164, 95

\bibitem[{{Strateva} {et~al.}(2001){Strateva}, {Ivezi{\'c}}, {Knapp},
  {Narayanan}, {Strauss}, {Gunn}, {Lupton}, {Schlegel}, {Bahcall}, {Brinkmann},
  {Brunner}, {Budav{\'a}ri}, {Csabai}, {Castander}, {Doi}, {Fukugita},
  {Gy{\H{o}}ry}, {Hamabe}, {Hennessy}, {Ichikawa}, {Kunszt}, {Lamb}, {McKay},
  {Okamura}, {Racusin}, {Sekiguchi}, {Schneider}, {Shimasaku}, \&
  {York}}]{Strateva2001}
{Strateva}, I., {Ivezi{\'c}}, {\v{Z}}., {Knapp}, G.~R., {et~al.} 2001, \aj,
  122, 1861

\bibitem[{{Tabatabaei} {et~al.}(2018){Tabatabaei}, {Minguez}, {Prieto}, \&
  {Fern{\'a}ndez-Ontiveros}}]{Tabatabaei2018}
{Tabatabaei}, F.~S., {Minguez}, P., {Prieto}, M.~A., \&
  {Fern{\'a}ndez-Ontiveros}, J.~A. 2018, Nature Astronomy, 2, 83

\bibitem[{{Tago} {et~al.}(2008){Tago}, {Einasto}, {Saar}, {Tempel}, {Einasto},
  {Vennik}, \& {M{\"u}ller}}]{Tago2008}
{Tago}, E., {Einasto}, J., {Saar}, E., {et~al.} 2008, \aap, 479, 927

\bibitem[{{Tempel} {et~al.}(2016){Tempel}, {Kipper}, {Tamm}, {Gramann},
  {Einasto}, {Sepp}, \& {Tuvikene}}]{Tempel2016a}
{Tempel}, E., {Kipper}, R., {Tamm}, A., {et~al.} 2016, \aap, 588, A14

\bibitem[{{Tempel} {et~al.}(2012){Tempel}, {Tago}, \&
  {Liivam{\"a}gi}}]{Tempel2012}
{Tempel}, E., {Tago}, E., \& {Liivam{\"a}gi}, L.~J. 2012, \aap, 540, A106

\bibitem[{{Tempel} {et~al.}(2017){Tempel}, {Tuvikene}, {Kipper}, \&
  {Libeskind}}]{Tempel2017}
{Tempel}, E., {Tuvikene}, T., {Kipper}, R., \& {Libeskind}, N.~I. 2017, \aap,
  602, A100

\bibitem[{{Tous} {et~al.}(2023){Tous}, {Dom{\'\i}nguez-S{\'a}nchez}, {Solanes},
  \& {Perea}}]{Tous2023}
{Tous}, J.~L., {Dom{\'\i}nguez-S{\'a}nchez}, H., {Solanes}, J.~M., \& {Perea},
  J.~D. 2023, \apj, 942, 48

\bibitem[{{Tutukov} \& {Fedorova}(2016)}]{Tutukov2016}
{Tutukov}, A.~V. \& {Fedorova}, A.~V. 2016, Astronomy Reports, 60, 116

\bibitem[{{van der Burg} {et~al.}(2017){van der Burg}, {Hoekstra}, {Muzzin},
  {Sif{\'o}n}, {Viola}, {Bremer}, {Brough}, {Driver}, {Erben}, {Heymans},
  {Hildebrandt}, {Holwerda}, {Klaes}, {Kuijken}, {McGee}, {Nakajima},
  {Napolitano}, {Norberg}, {Taylor}, \& {Valentijn}}]{vanderBurg2017}
{van der Burg}, R. F.~J., {Hoekstra}, H., {Muzzin}, A., {et~al.} 2017, \aap,
  607, A79

\bibitem[{{Wilman} \& {Erwin}(2012)}]{Wilman2012}
{Wilman}, D.~J. \& {Erwin}, P. 2012, \apj, 746, 160

\bibitem[{{Yang} {et~al.}(2022){Yang}, {Irwin}, {Li}, {Wiegert}, {Wang}, {Sun},
  {Damas-Segovia}, {Li}, {Shen}, {Walterbos}, \& {Vargas}}]{Yang2022}
{Yang}, Y., {Irwin}, J., {Li}, J., {et~al.} 2022, \apj, 927, 4

\end{thebibliography}

\end{document}